\documentclass[12pt]{iopart}
\usepackage{iopams}  
\usepackage{graphicx}
\DeclareMathAlphabet{\mathpzc}{OT1}{pzc}{m}{it}
\begin{document}

\title[Variable placement of templates technique for binary inspiral searches]{Variable placement of templates
technique in a 2D parameter space for binary inspiral searches}

\author{
Fabrice Beauville,
Damir Buskulic
\footnote[3]{Corresponding author, email: buskulic@lapp.in2p3.fr},
Raffaele~Flaminio,
Romain~Gouaty,
Daniel~Grosjean,
Fr\'ed\'erique~Marion,
Benoit~Mours,
Edwige~Tournefier,
Didier~Verkindt,
Michel~Yvert
}

\address{LAPP/Universit\'e de Savoie, Chemin de Bellevue, B.P. 110,
74941 Annecy-le-Vieux Cedex, FRANCE}

\begin{abstract}
In the search for binary systems inspiral signal in interferometric
gravitational waves detectors, one needs the generation and placement
of a grid of templates. We present an original technique for the
placement in the associated parameter space, that
makes use of the variation of size of the isomatch ellipses
in order to reduce the number of templates necessary
to cover the parameter space. This technique avoids the potentially
expensive computation of the metric at every point, at the cost of
having a small number of ``holes'' in the coverage, representing
a few percent of the surface of the parameter space, where
the match is slightly lower than specified.
A study of the covering efficiency, as well as a comparison
with a very simple regular tiling using a single ellipse is made.
Simulations show an improvement varying between 6\% and 30\%
for the computing cost in this comparison.
\end{abstract}



\maketitle

\section{Introduction}
In searching for gravitational wave signals from coalescing binary compact objects,
one commonly uses an optimal filtering technique \cite{owen}. This technique consists
of the comparison of the output signal of an interferometric gravitational waves
detector with a family of expected theoretical waveforms, called templates. Each template
depends on one or more parameters $\left\{\lambda_i\right\}$.
The choice of the templates in the $\left\{\lambda_i\right\}$ parameter space,
called placement, is the purpose of this paper. We restrict ourselves to
a 2D parameter space, considering spinless templates computed at second post-newtonian order.

\par We will first describe in section \ref{sec_motivations} the motivations of our
placement technique, comparing it with a simple uniform paving of the parameter space.
Section \ref{sec_computation_par} describes the calculation of the parameters of the
parameter space portion covered by a single template. This portion is
in our case well approximated by an ellipse.
Next, section \ref{sec_triangulation} treats the triangulation of the parameter
space, a step needed by the placement, which is covered by section \ref{sec_placement}.
Finally, performance tests are covered by section \ref{sec_performance}, where
some real use-cases are considered in the context of the Virgo detector \cite{virgo}.

\section{Motivations}
\label{sec_motivations}

\subsection{Portion of the parameter space covered by one template}

The comparison of a signal with one template is made through a Wiener filter \cite{wiener}:
\begin{equation} \label{optimal_filter_eqn}
\langle \tilde{a},\widetilde{T} \rangle = 2 \left[
\int^{f_s}_{f_i} \frac{\tilde{a}(f).\widetilde{T}^*(f)}{S(f)}\textrm{d} f
+ \textrm{c.c.}\right]
\end{equation}
This is essentially a weighted intercorrelation, $\tilde{a}(f)$ being
the interferometer output and $\widetilde{T}(f)$ the template. $S(f)$ is
the noise power spectral density (PSD) of the detector, $f_i$ and $f_s$
are the lower and upper limits of the detector spectral window.
\par Each template is represented by a point in a multidimensional parameter
space. After taking care of most extrinsic parameters (like time of arrival or
initial orbital phase of the system) by maximizing the output
of the optimal filter over them \cite{schutz}, there remain only two parameters,
that we will call $\lambda_1$ and $\lambda_2$. Those parameters may be
the masses of the two bodies but in general, one uses parameters derived from the
masses that simplify the calculations.
\par A template corresponding to parameters $(\lambda_1,\lambda_2)$ is
sensitive to a signal corresponding to nearby parameters
$(\lambda_1+\delta\lambda_1 ,\lambda_2+\delta\lambda_2)$. The difference leads to
a decrease in signal over noise ratio (SNR) with respect to the SNR obtained
with a signal corresponding to the exact template. For an acceptable loss in SNR,
each template covers a portion of the two dimensional parameter space.
Following Owen \cite{owen} in a geometrical interpretation of the optimal
filtering, one is able to define a distance between two templates as the
ambiguity function maximized over extrinsic parameters,
called ``match''. When filtering
a signal which has the same shape as a template of parameters
$(\lambda_1+\delta\lambda_1 ,\lambda_2+\delta\lambda_2)$
with a reference template of parameters $(\lambda_1,\lambda_2)$,
the match is the fraction of the optimal SNR
obtained when filtering the reference template with a signal identical in shape
to itself. 

\par Given a minimal match $MM$, we can define the region of parameter
space around a given point corresponding to a template $T$, the match of which,
computed with any template corresponding to a point in the region,
will be above $MM$.
We will call the boundary of this region the ``isomatch
contour''. The shape of this boundary may be complex, so one generally uses
parameters for which it has been shown that, for high values of the minimal match,
($MM>0.97$) the contour is closed and well approximated by an ellipse
\cite{owen}. Throughout this paper,instead of masses,
we will use chirp times $\tau_0$ and $\tau_{1.5} $\cite{chirp_times_def}
defined as:

\begin{equation}
\tau_0=\frac{5M}{256\eta (\pi Mf_0)^{8/3}}
\qquad
\tau_{1.5}=\frac{\pi M}{8\eta (\pi Mf_0)^{5/3}}
\end{equation}

in geometrized units ($G=c=1$), where $M$ is the total mass of the binary
system, $\eta=m_1m_2/M^2$ is the
symmetric mass ratio and $f_0$ a fiducial frequency chosen as the lower
frequency cutoff of the detector sensitivity. Results are properly scaled
to restore physical units.
\par The calculation of the parameters of the ellipse may be done
analytically for a given spectral density \cite{owen}\cite{lal_calcul_ellipse}.

\par The final goal of our study is to pave the parameter space with isomatch contours
in as optimal a way as possible. This is equivalent to finding the minimal
set of templates whose isomatch contours pave all the parameter space,
without letting any hole or unpaved region \cite{pinto}.

\subsection{Simple paving of the parameter space}
\label{simple_pave}
\begin{figure}[htb]
\begin{center}
\includegraphics[width=150mm]{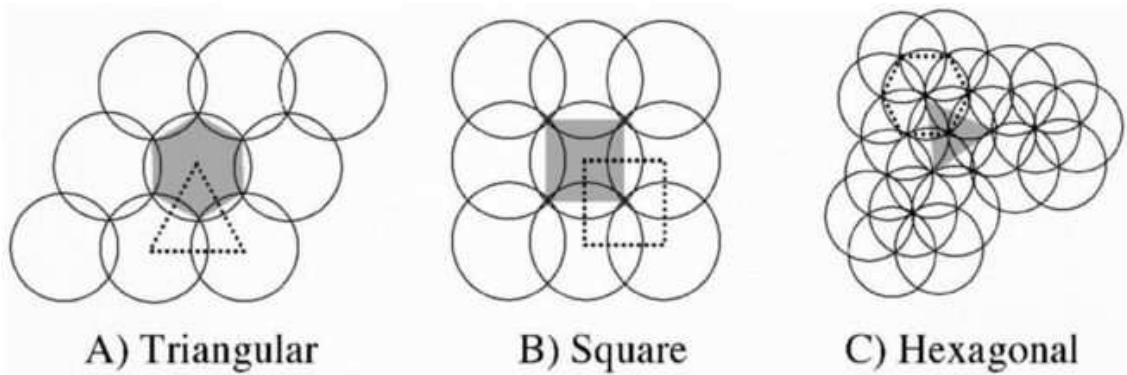}
\end{center}
\caption{\label{circle_place}\footnotesize{Paving of a plane with different
elementary cells. The relevant Voronoi sets are shown in gray.}}
\end{figure}

One simple solution, already described elsewhere, is to calculate
the ellipse parameters for the point in the parameter space where it is known to be
the smallest and pave the space with this single ellipse \cite{owen},
obtaining a regular tiling of the parameter space.
This is not very different from paving a bidimensional space with circles.
As was already noted \cite{pinto},
because of the rotational symmetry, the centers of the circles should
sit at the vertices of regular polygons which make a regular tiling
of the plane. This is only possible for triangles, squares or hexagons.
In the first case, the centers of the circles are placed on the corners
of an equilateral triangle, as shown in figure \ref{circle_place} A).
It is desirable to have the sparsest possible circles, which means that
three circles touch at one single point $P$. The surface region
consisting of the points whose closest circle center is $C$
is shown in gray. This is also the surface covered on average by one circle.
In the triangular case, it is a hexagon. The set of points which belongs
to this region is called the Voronoi set of $C$. As illustrated in
figure \ref{circle_place}, in the case of a
square tiling, the Voronoi set has a square shape and in the case of
a hexagonal tiling, the Voronoi set has a triangular shape.
It has been shown \cite{pinto}, as one would intuitively expect,
that the most efficient tiling in the case of placement of circles is
the triangular one.
Of course, in our case, the circles are skewed according to the parameters
of the initially calculated ellipse.

\par
\begin{figure}[htb]
\begin{center}
\includegraphics[width=130mm]{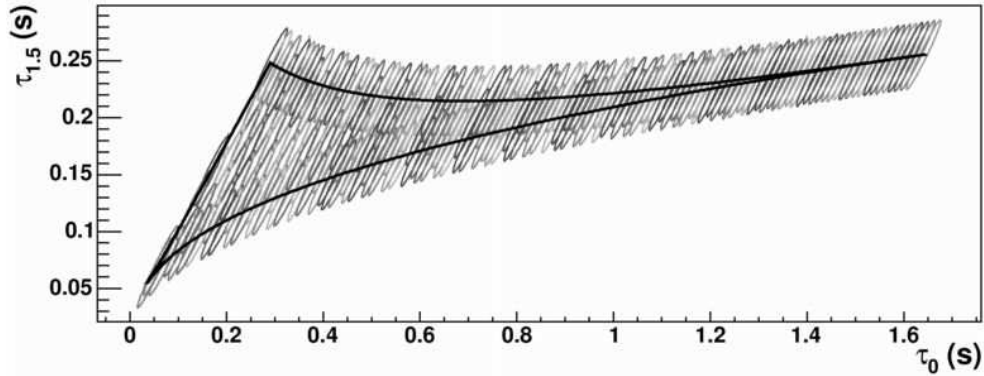}
\end{center}
\caption{\label{simple_tile}\footnotesize{Example of a regular tiling of
a parameter space. The templates are computed at 2 PN order in the mass range
[5;50] $\textrm{M}_\odot$,
the minimal match being 0.95, the frequency range [50;2000] Hz with the Virgo PSD.}}
\end{figure}

 The tiling is extended outside the parameter space to make
the coverage complete.
The ellipses, the center of which lies in a physically forbidden region (under the
equal mass line), are shifted towards the allowed region, staying on the
equal mass line, still ensuring the completeness of the coverage.
An example is given in fig. \ref{simple_tile}, where the ellipse at the extreme
right (smallest masses) represents the only computed point.

\subsection{Improvements to this method}
The above simple method is very fast but,
assuming that one uses the smallest possible ellipse,
is clearly suboptimal in most cases. It gives a higher number of templates
than would be ideally needed if one was able to calculate
the shape of the isomatch contours at any given point of the parameter space
and use those bigger shapes to cover the space.
A second problem would then arise, since an optimal tiling of the parameter
space with varying shapes is far from being obvious. The principle of
reconstruction of exact isomatch contours has been described previously
\cite{buskulic_gwdaw02} as well as a preliminary placement method.
\par We present in this paper an extension and improvement of this method in the case
where the elliptic approximation for isomatch contours is assumed valid.

\section{Computation of ellipse parameters}
\label{sec_computation_par}
Before doing the placement, one should be able to calculate
as fast as possible the ellipse parameters at any given point
in the parameter space. This is done by
\begin{itemize}
\item Calculating the ellipses at a chosen set of points
(we obtain ``seed ellipses'').
\item Triangulating the parameter space with this set.
Actually, as we will see, those two steps are closely linked. We
give in appendix a short tutorial about triangulation and computational geometry.
\item Interpolate linearly ellipses at any point using the previously calculated
seed ellipses. This step is much faster than an analytical computation.
\end{itemize}
\subsection{Computation of seed ellipses}
The seed ellipses are computed using the algorithm included
inside the LIGO Analysis Library (LAL) \cite{lal}.
This algorithm uses the procedure described in \cite{owen_2}. The metric
components used to find the parameters of the ellipse are calculated
using the moments of the PSD curve.

\subsection{Triangulation and interpolation}
\begin{figure}[htb]
\begin{center}
\includegraphics[width=120mm]{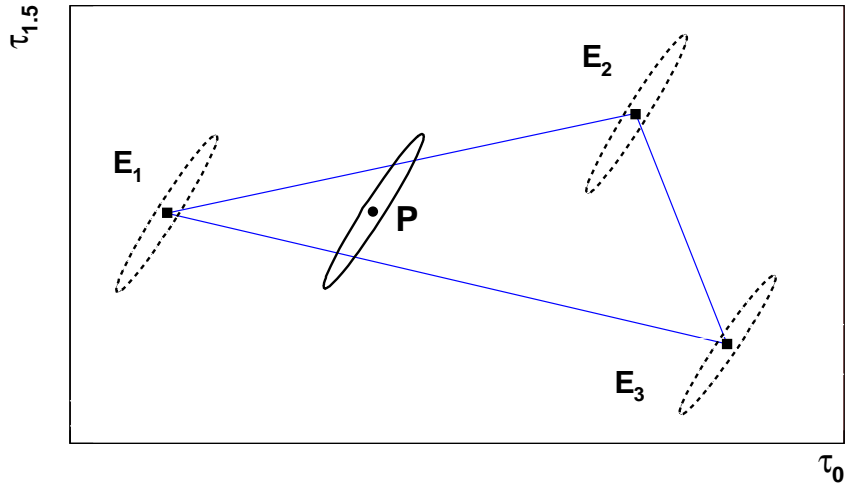}
\end{center}
\caption{\label{interpolation}\footnotesize{Linear interpolation between analytically
calculated ellipses $(E_1, E_2, E_3)$, that may differ in size and orientation,
at an arbitrary point P in a triangle.}}
\end{figure}

The triangulation of the parameter space deserves hereafter a section by itself.
Once it is computed, each point $P$ in the parameter space belongs to one and only one
triangle whose corners are three seed points. One is able to interpolate
linearly the shapes (resp. metric parameters) of the three seed ellipses
to obtain the parameters of the ellipse (resp. metric) at point $P$
(see fig. \ref{interpolation}).
\section{Triangulation of the parameter space}
\label{sec_triangulation}
The triangulation of the parameter space is done using standard techniques known
in computational geometry. The notions necessary to understand
the present study are explained in appendix. The base algorithm
used is known as the Bowyer-Watson \cite{watson}\cite{bowyer} algorithm.

\subsection{Triangulation algorithm adapted to the CB parameter space}
The Bowyer-Watson algorithm is quite simple but needs adaptation to our problem.
We need to take care of the fact that the borders of the parameter space
are not convex and we need to choose which points to use for the triangulation.
\par
\begin{figure}[htb]
\begin{center}
\includegraphics[height=35mm]{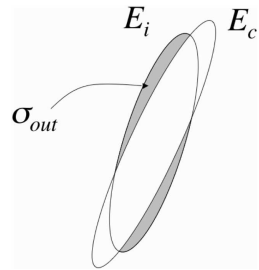}
\end{center}
\caption{\label{intersect_ellipses}\footnotesize{ The variable indicating the difference between
the two interpolated and calculated ellipses at the same point is the proportion
of the surface of the interpolated ellipse not common to both ellipses.}}
\end{figure}

 The main idea of our adapted algorithm is to start from an existing triangle
at the corners of which sit three already calculated ellipses $\{E_1, E_2, E_3\}$
and subdivide it only if necessary, i.e. if for any point $P$ inside the triangle,
the ellipse linearly interpolated between $\{E_1, E_2, E_3\}$
is different enough from the one calculated using the metric at that point. 
Let $E_i$ be the interpolated ellipse and $E_c$ the calculated one.
$\sigma$ being the measure of the surface of $E_i$,
$\sigma_{out}$ the surface of $E_i$ that does not intersect $E_c$
(fig. \ref{intersect_ellipses}), the variable describing the difference
between $E_i$ and $E_c$ has been chosen as the proportion
\begin{equation} \label{proportion_limit}
p = \frac{\sigma_{out}}{\sigma}
\end{equation}
It was not deemed necessary to also take into account the surface
of $E_c$ that does not intersect $E_i$, because if $\sigma_{out}$
is null, the interpolated ellipse is completely inscribed inside
the calculated one and we are simply going to make a more dense
placement at a later stage.
A limit is set on this variable to stop the subdivision of triangles.

   \subsubsection{Division of an existing triangle}
Given an existing triangle, a choice has to be made on the points appropriate for
its subdivision. Ideally, one would use the points which have the highest proportion
$p$. It is however impractical, and very expensive in terms of computing power to test
all the points in a triangle to find the one with the higher $p$. We chose to
test only the middle points of each segment forming the triangle.
\par
\begin{figure}[htb]
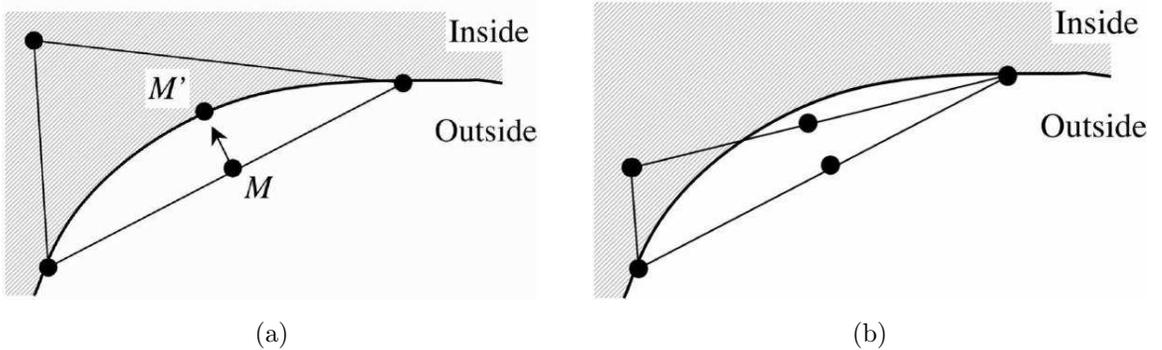

\begin{center}
\begin{tabular}{cc}
\mbox{
\includegraphics[height=40mm]{close_border_point.epsf}
} &
\mbox{
\includegraphics[height=40mm]{close_border_peculiar.epsf}
} \\
\mbox{
\footnotesize{(a)}
} &
\mbox{
\footnotesize{(b)}
}
\end{tabular}
\end{center}
\caption{\label{close_border_point}\footnotesize{ Replacement of a point $M$ outside the parameter
space by a point $M'$ on the border close by (a). A peculiar case, which is taken
into account, is also shown (b)}}
\end{figure}

 Each of these three points is inserted and used to subdivide the triangle
following a Delaunay method, but considering
only the triangle, not the adjacent ones that may exist in the ongoing
triangulation process. If the middle point of a segment is outside the parameter space,
it is replaced by the closest point on the border, perpendicularly to the segment
(fig. \ref{close_border_point}). Some peculiar situations (two middle segment
points outside the parameter space for example) are taken into account. All subtriangles
generated outside the parameter space are removed.
\par From the description above, it is obvious that the final triangulation
will not be strictly speaking a Delaunay one, since we use the Delaunay criterion
only locally for a triangle subdivision.

   \subsubsection{Global algorithm view}
   \label{global_view}
We start from the triangle formed in the $(\tau_0,\tau_{1.5})$ parameter space by
the three angular points corresponding to $(m_{min},m_{min})$, $(m_{min},m_{max})$
and $(m_{max},m_{max})$, where $m_{min}$ and $m_{max}$ are respectively the
minimal and maximal masses of the binary system members considered.
\par
\begin{figure}[!htb]
\begin{center}
\begin{tabular}{cc}
\mbox{
\includegraphics[width=70mm]{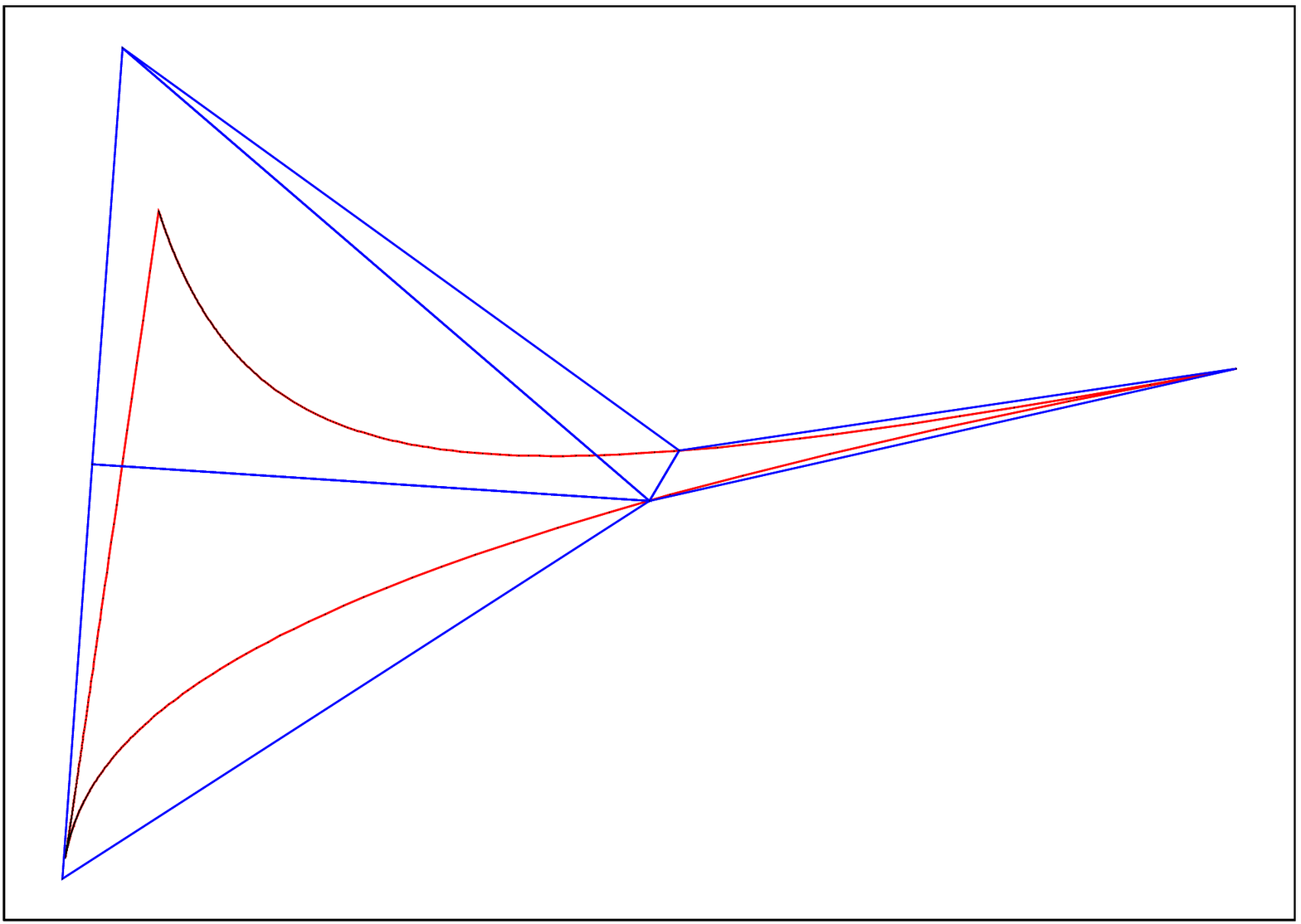}
} &
\mbox{
\includegraphics[width=70mm]{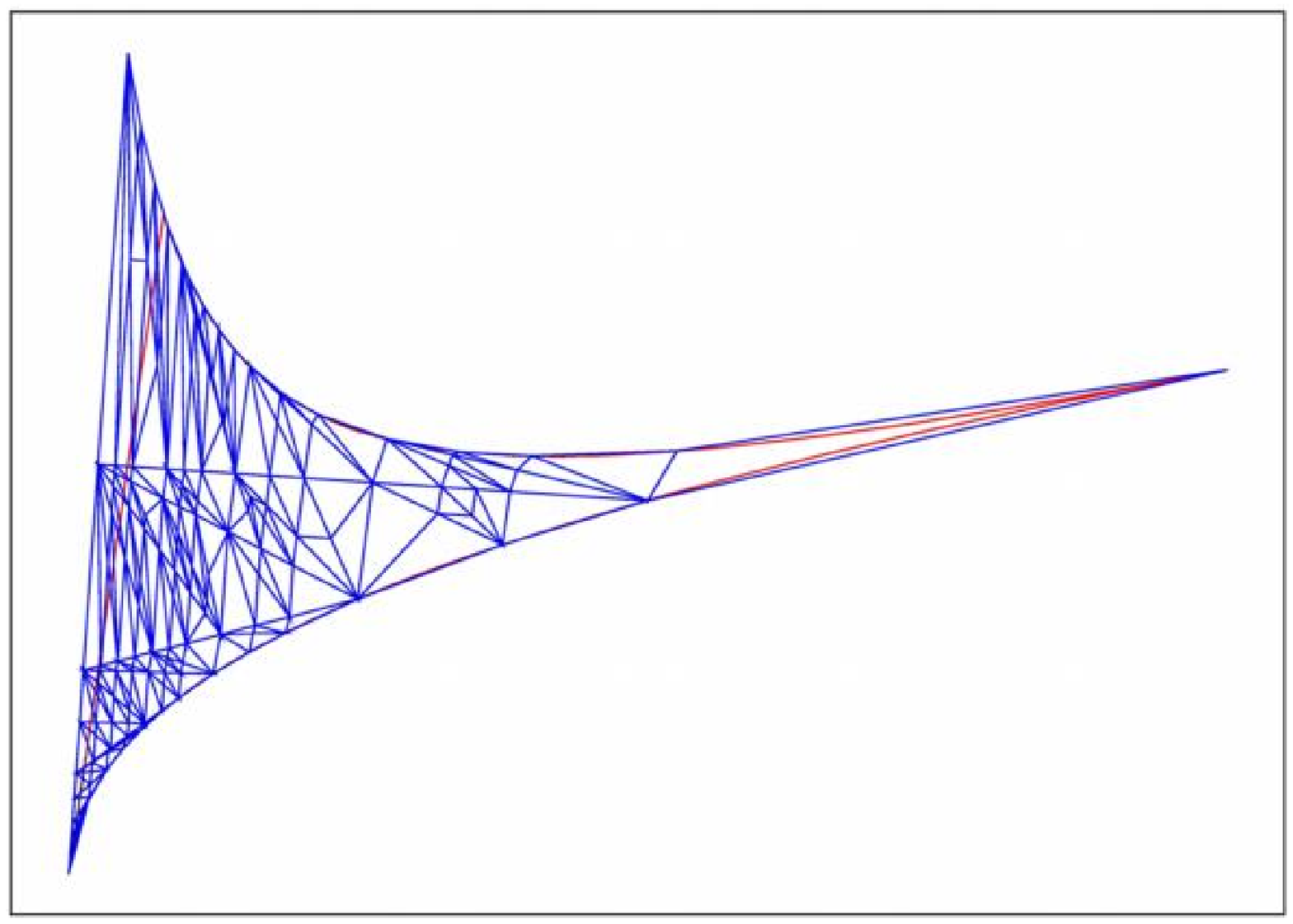}
} \\
\mbox{
\footnotesize{Refinement step 1}
} &
\mbox{
\footnotesize{Refinement step 5}
}
\end{tabular}
\end{center}
\caption{\label{triangulation_steps}\footnotesize{Mesh refinement steps. Starting from
a triangle enclosing the parameter space, insert points and retriangulate
while the proportion of discrepancy $p$ between interpolated and calculated
ellipse is greater than a limit $p_{lim}$.}}
\end{figure}

 The triangle is then recursively subdivided as described above. Since there is
a limit on the $p$ proportion of each inserted point, the subdivision will stop
naturally when the mesh becomes dense enough. These successive
refinement steps are illustrated in figure \ref{triangulation_steps}.
In order to avoid too big a number of calculated ellipses, and to
limit the computing time, the number
of refinement steps has been arbitrarily limited to 7. It was verified that this
doesn't bring any problems, except in the lower left corner of the parameter space,
corresponding to high masses (above 10 $\textrm{M}_\odot$) for both objects.
In that case, the placement may be somewhat wrong but a posteriori Monte-Carlo tests
show an undercoverage not exceeding 2\% of the total parameter space surface.

As may be noted on figure \ref{triangulation_steps}, the tesselation of the parameter
space is extended in the physically allowed region to avoid some extrapolation side-effects
in the following placement procedure.

\subsection{Extrapolation outside border of the parameter space}
Each ellipse calculated for the placement procedure described hereafter is actually
interpolated inside one of the triangles found during the triangulation step.
If the point considered by the placement is outside of the tesselated (triangulated) part,
it doesn't belong to any triangle a priori. We will see that the placement procedure needs
to spill over the strict borders of the parameter space to ensure complete coverage,
and it may happen that a determination of ellipse parameters is needed outside
the tesselated part.
Furthermore, the calculation of the metric is impossible in the disallowed (physically forbidden)
region under the equal mass line in the $(\tau_0,\tau_{1.5})$ parameter space. Therefore,
we cannot triangulate that region since we cannot calculate true ellipses or contours in it.
\par Thus, we need to provide a way to extrapolate the ellipse parameters outside the strictly
tesselated part of the space. As will be seen later, the final step of the
placement procedure consists of shifting the points found in the forbidden region
so that they fall in the allowed one. But extrapolation is needed all around
the space border during the placement, albeit in a limited area.
\par
\begin{figure}[!htb]
\begin{center}
\includegraphics[width=90mm]{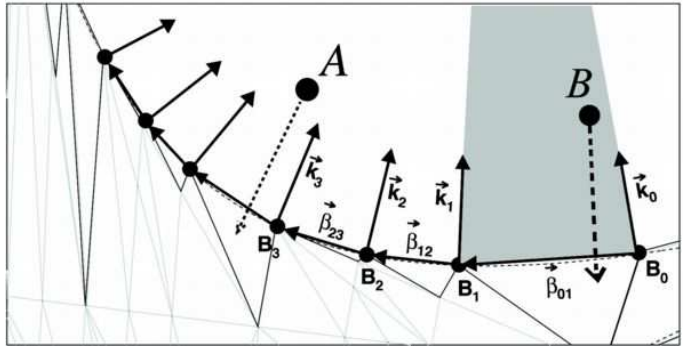}
\end{center}
\caption{\label{associate_triang_border}\footnotesize{Association between a point outside the
triangulated part of the parameter space and a triangle on the border}}
\end{figure}

 For a given point $A$ outside the parameter space, it is natural to associate it with the
closest triangle of the tesselation. The word "closest" should be taken with care, as
closest in euclidian distance doesn't mean more adequate for our purposes.
We consider only the triangles which border the tesselated region, i.e. which have
one side that is not common to another triangle, thus delimits the border of the
tesselation (fig. \ref{associate_triang_border}). Once the triangle associated
with the point $A$ is determined, one can do a linear extrapolation of the ellipse
parameters, as for the points inside the triangle.

The choice of the triangle associated with a given point $A$ is done as follows. We define the
vectors $\vec \beta_{ij}$ which join two successive vertices $B_i$ and $B_j$ lying on the 
border of the tesselated part of the parameter space.
Each vertex $B_i$ is associated with a vector $\vec k_i$
whose direction is pointing towards the outside of the space and is an
average of the normal to two consecutive vectors $\vec \beta_{ki}$
and $\vec \beta_{ij}$.
\par A point $A$ will be associated with the triangle containing the vertices $B_i$ and $B_j$
if it is located in the domain delimited by $\vec \beta_{ij}$ and the two lines
defined by $(B_i,\vec k_i)$ and $(B_j,\vec k_j)$. An example of such a domain is shown
in gray in figure \ref{associate_triang_border}.
\par Clearly, the very simple extrapolation we describe is valid only for the points close
to the space boundary. The lines $(B_i,\vec k_i)$ will cross and it is not possible
to associate a point and a triangle beyond those crossings. Furthermore,
the validity of the extrapolation is not guaranteed for points pushed away
from the boundary of the parameter space.
In our case, where we marginally extend the calculation of the metric outside the borders,
this shows not to be a problem.

\subsection{Results in concrete cases}
\begin{figure}[!htb]
\begin{center}
\begin{tabular}{cc}
\mbox{
\includegraphics[height=50mm]{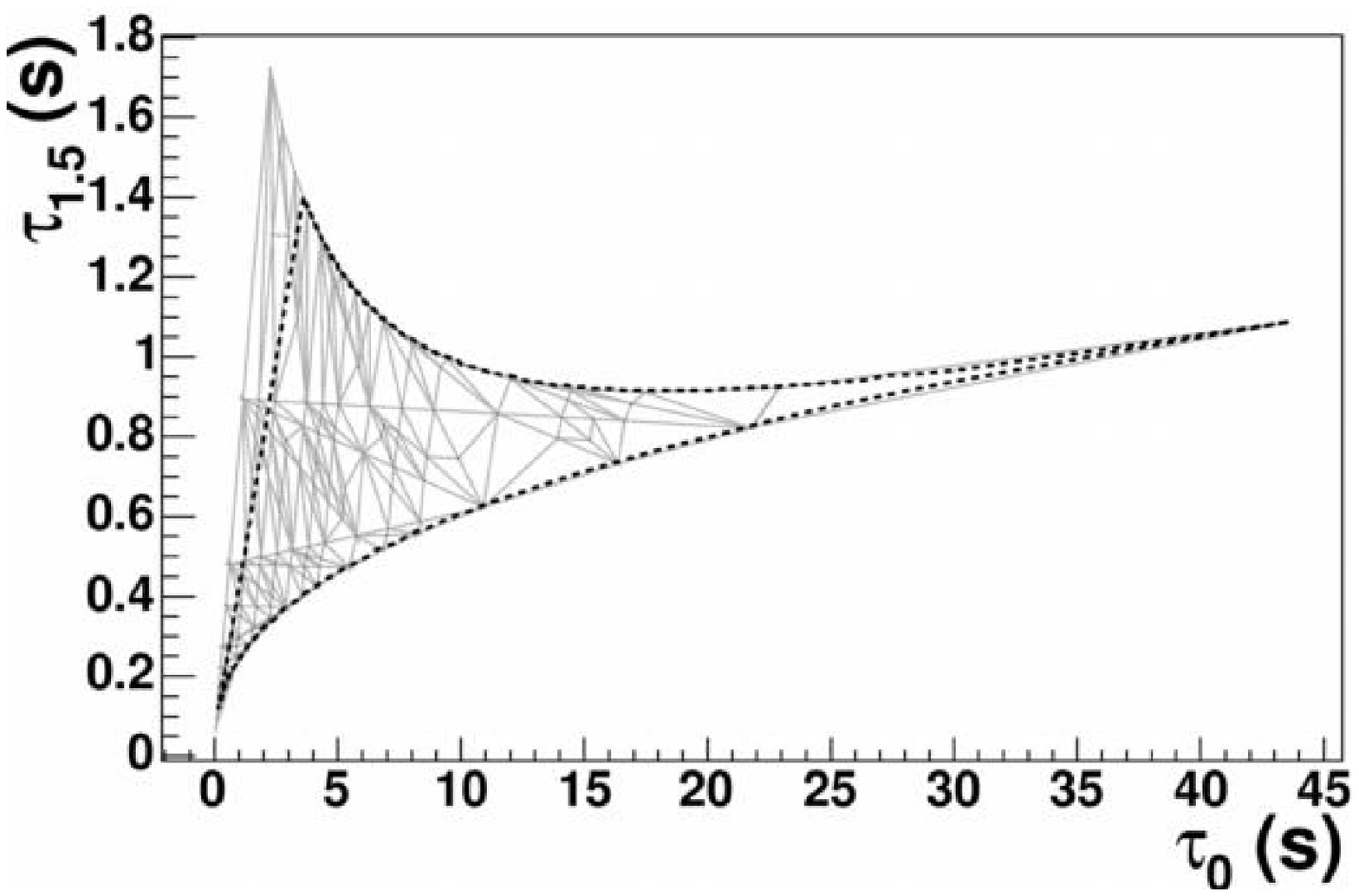}
} &
\mbox{
\includegraphics[height=50mm]{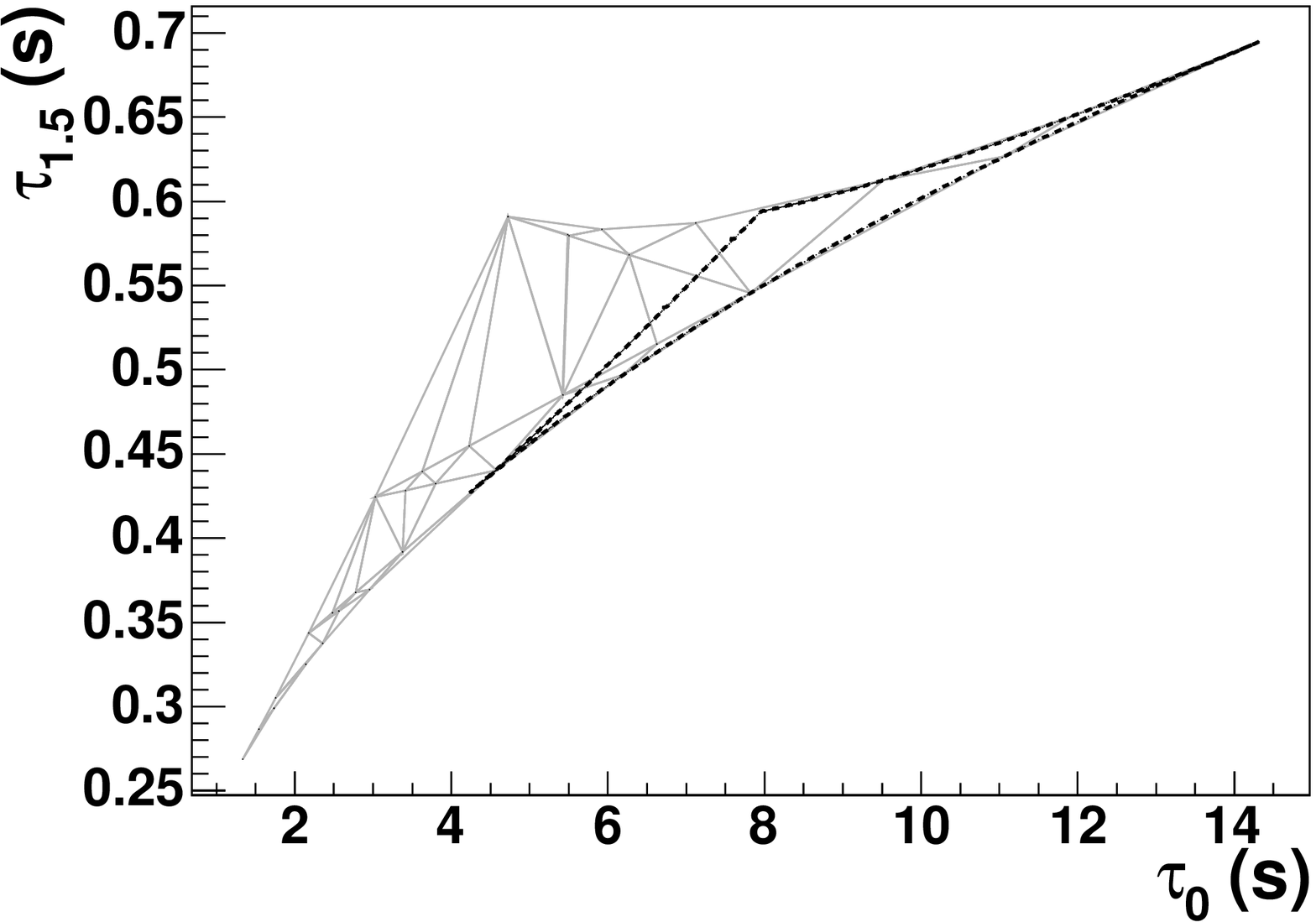}
} \\
\mbox{
\footnotesize{$m_{min}=1 \textrm{M}_\odot$, $m_{max}=30 \textrm{M}_\odot$,}
} &
\mbox{
\footnotesize{$m_{min}=1.95 \textrm{M}_\odot$, $m_{max}=4.05 \textrm{M}_\odot$,}
} \\
\mbox{
\footnotesize{$F_l=40$ Hz, $F_h=2000$ Hz,}
} &
\mbox{
\footnotesize{$F_l=40$ Hz, $F_h=2000$ Hz,}
} \\
\mbox{
\footnotesize{$PN=2$, $NStep_{max}=5$, $N_T=152$}
} &
\mbox{
\footnotesize{$PN=2$, $NStep_{final}=4$, $N_T=31$}
}
\end{tabular}
\end{center}
\caption{\label{concrete_triang}\footnotesize{Examples of triangulations
representing real use cases for CB searches in Virgo. The black line represents
the border of the parameter space. The triangulation area is extended
outside to prevent extrapolation problems.
See the text for an explanation of computing conditions.}}
\end{figure}

Figure \ref{concrete_triang} shows triangulation in a few real cases.

\begin{figure}[htb]
\begin{center}
\includegraphics[width=100mm]{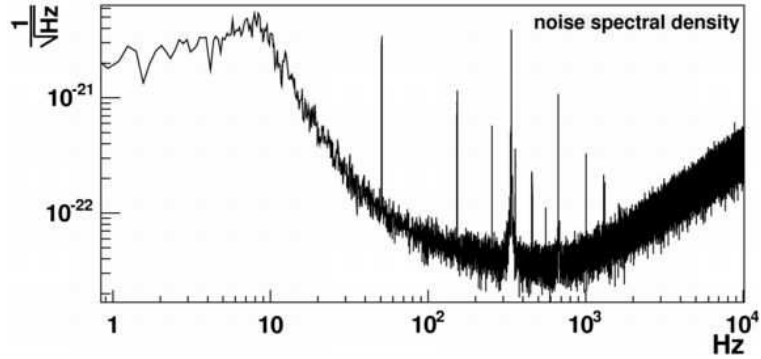}
\end{center}
\caption{\label{spectrum_triang}Noise spectral density for the real use cases.}
\end{figure}

\begin{itemize}
\item $m_{min}$ and $m_{max}$ are the minimal and maximal masses of the parameter
space
\item $F_l$ and $F_h$ the lower and higher frequency cutoffs used for the
generation of templates
\item $PN$ is the order of the post-newtonian expansion
\item $NStep_{max}$ is the limit imposed on the number of triangulation steps
\item $NStep_{final}$ is the number of steps effectively needed to satisfy
the surface proportion condition for all the ellipses generated, without
reaching the $NStep_{max}$ limit
\item $N_T$ is the number of calculated points in the triangulation to reach the
$NStep_{max}$ or $NStep_{final}$ limit
\item The noise spectral density used was a Virgo-like one, shown on fig. \ref{spectrum_triang}
\end{itemize}

\section{Placement}
\label{sec_placement}
   \subsection{Isomatch properties}
\label{isomatch_properties}
\begin{figure}[htb]
   \begin{minipage}[t]{0.45\linewidth}
      \begin{center}
         \includegraphics[height=55mm]{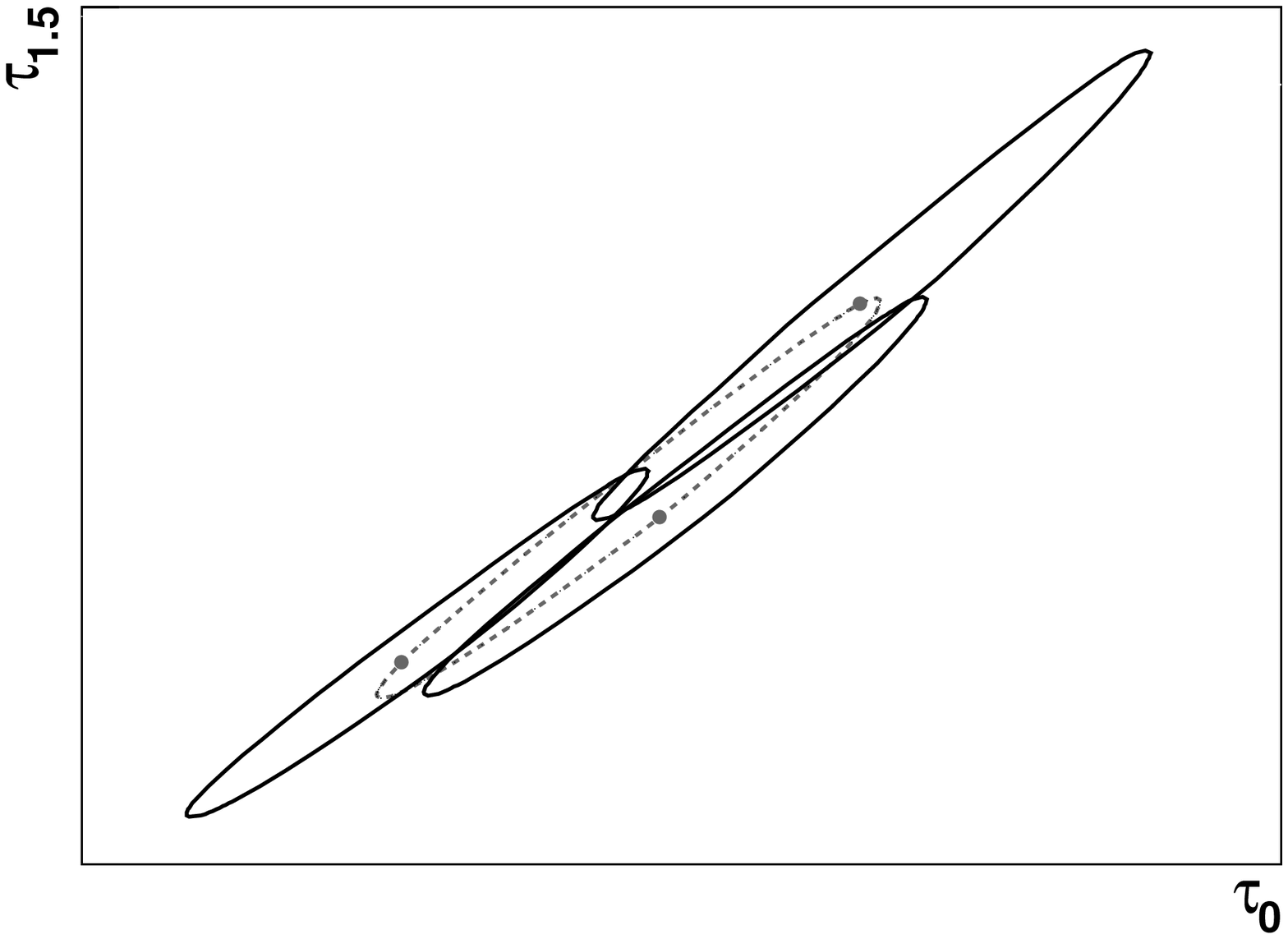}
      \end{center}
      \caption{\label{placement_three}\footnotesize{Placement of three contours in the 
($\tau_0$,$\tau_{1.5}$) plane}}
   \end{minipage}
   \hspace{0.1\linewidth}
   \begin{minipage}[t]{0.45\linewidth}
      \begin{center}
         \includegraphics[height=55mm]{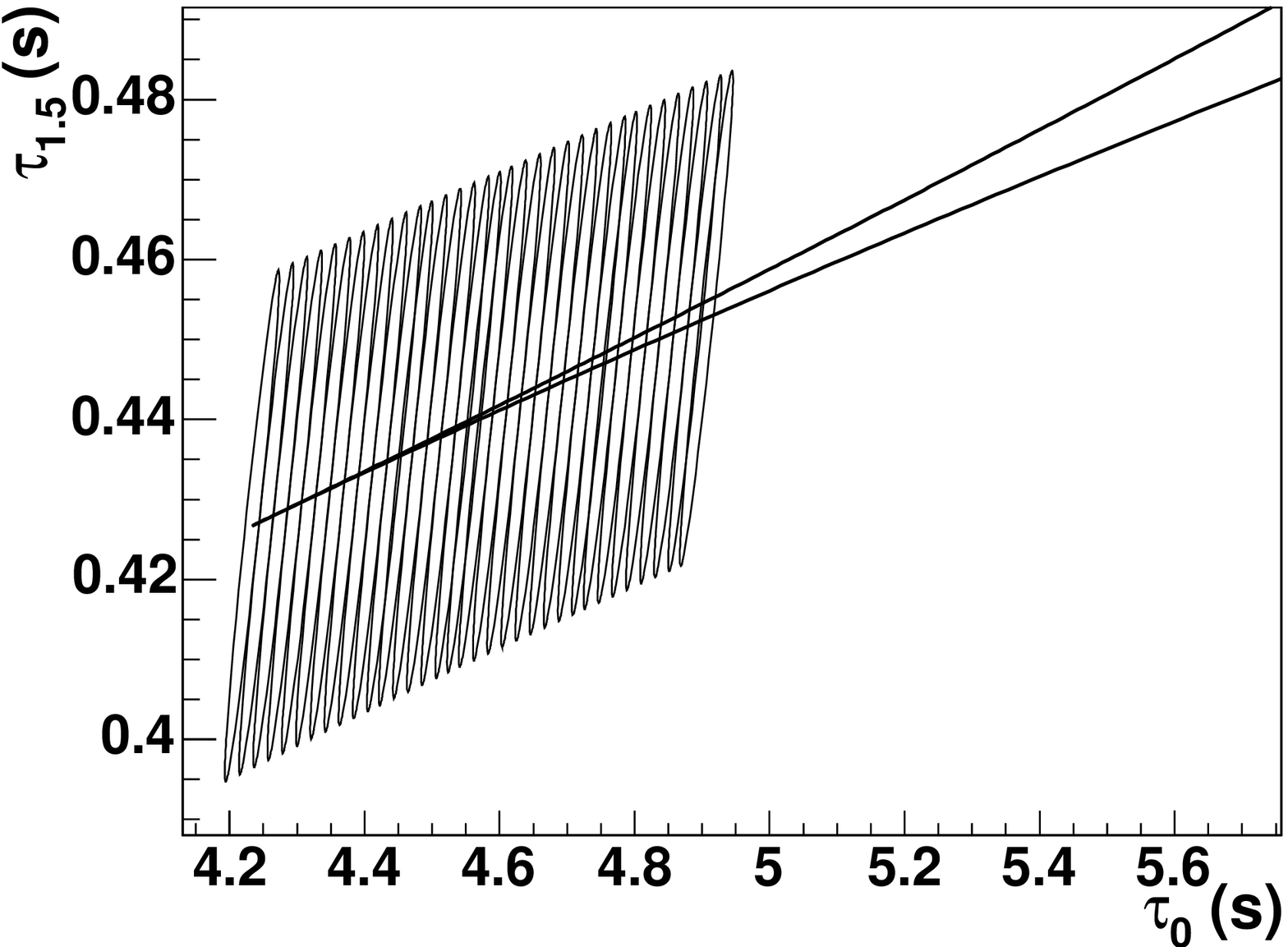}
      \end{center}
      \caption{\label{placement_stage1}\footnotesize{First stage of placement along the
equal mass line}}
   \end{minipage}
\end{figure}

Once the triangulation and seed ellipses have been generated, the placement
is done in two stages. Both rely on basic properties of isomatch contours
described in \cite{buskulic_gwdaw02}, namely:
\begin{itemize}
\item The match symmetry between two contours. If $\widetilde{T}_1$
and $\widetilde{T}_2$ are two normalized templates, one has 
$\langle \widetilde{T}_1,\widetilde{T}_2 \rangle =
\langle \widetilde{T}_2,\widetilde{T}_1 \rangle = M$.
Thus, the point in the parameter space corresponding to $\widetilde{T}_1$ is
located on the isomatch contour of value $M$ corresponding to $\widetilde{T}_2$,
and conversely, the point corresponding to $\widetilde{T}_2$ is
located on the isomatch contour of value $M$ corresponding to $\widetilde{T}_1$.
In practical computations, the match symmetry may not be absolutely verified
because in general one maximizes over the initial phase of one template (say $\widetilde{T}_1$), which is
not done for the signal ($\widetilde{T}_2$). This has proven to be negligible
for smooth variations of the metric throughout the parameter space, which is
roughly the case in our tests using the LAL, except perhaps for high masses,
$>10 \textrm{ M}_{\odot}$.
\item To place an ellipse with respect to another in an optimal way, one
introduces a guiding ellipse. This allows to place three ellipse sets
(fig. \ref{placement_three}). The three ellipses intersect at the center of
the guiding ellipse.
\end{itemize}

In the course of the running of the algorithm,
if two of the ellipses are placed, the third one may be positioned
naturally on the border of the guiding ellipse by maximizing the
surface of the triangle formed by the centers of the three ellipses.

   \subsection{First stage}
The first stage consists of a side by side placement
along the equal mass line starting from the $(m_{max}, m_{max})$ point
(fig. \ref{placement_stage1}). Unlike in the simple placement case where this was avoided,
it is the most efficient way of paving while only one ellipse is needed to cover
the parameter space along the direction of the semi-major axis of the ellipse, an almost vertical
direction in most of our cases.
\par
\begin{figure}[htb]
\begin{center}
\begin{tabular}{cc}
\mbox{
\includegraphics[height=65mm]{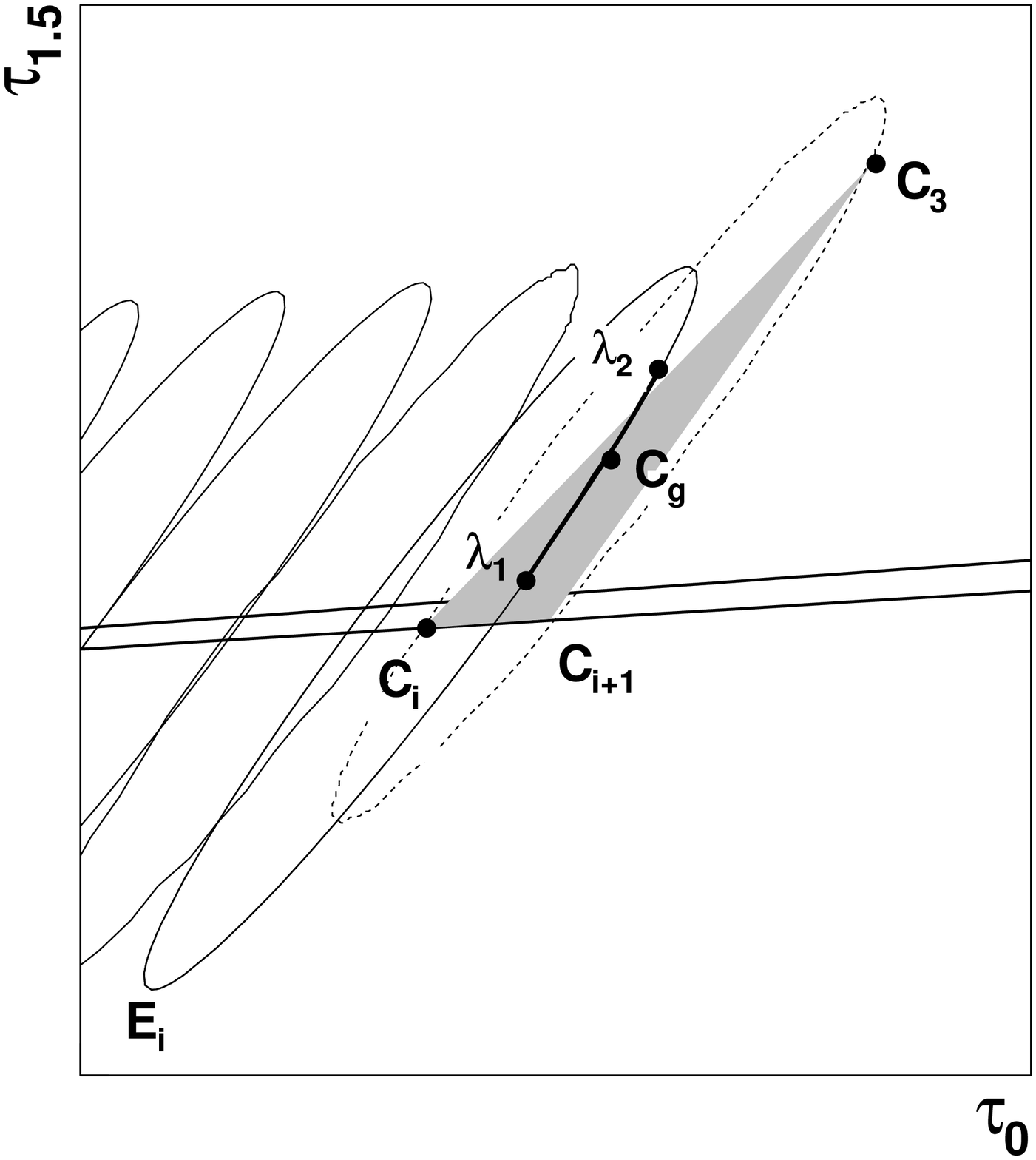}
} &
\mbox{
\includegraphics[height=65mm]{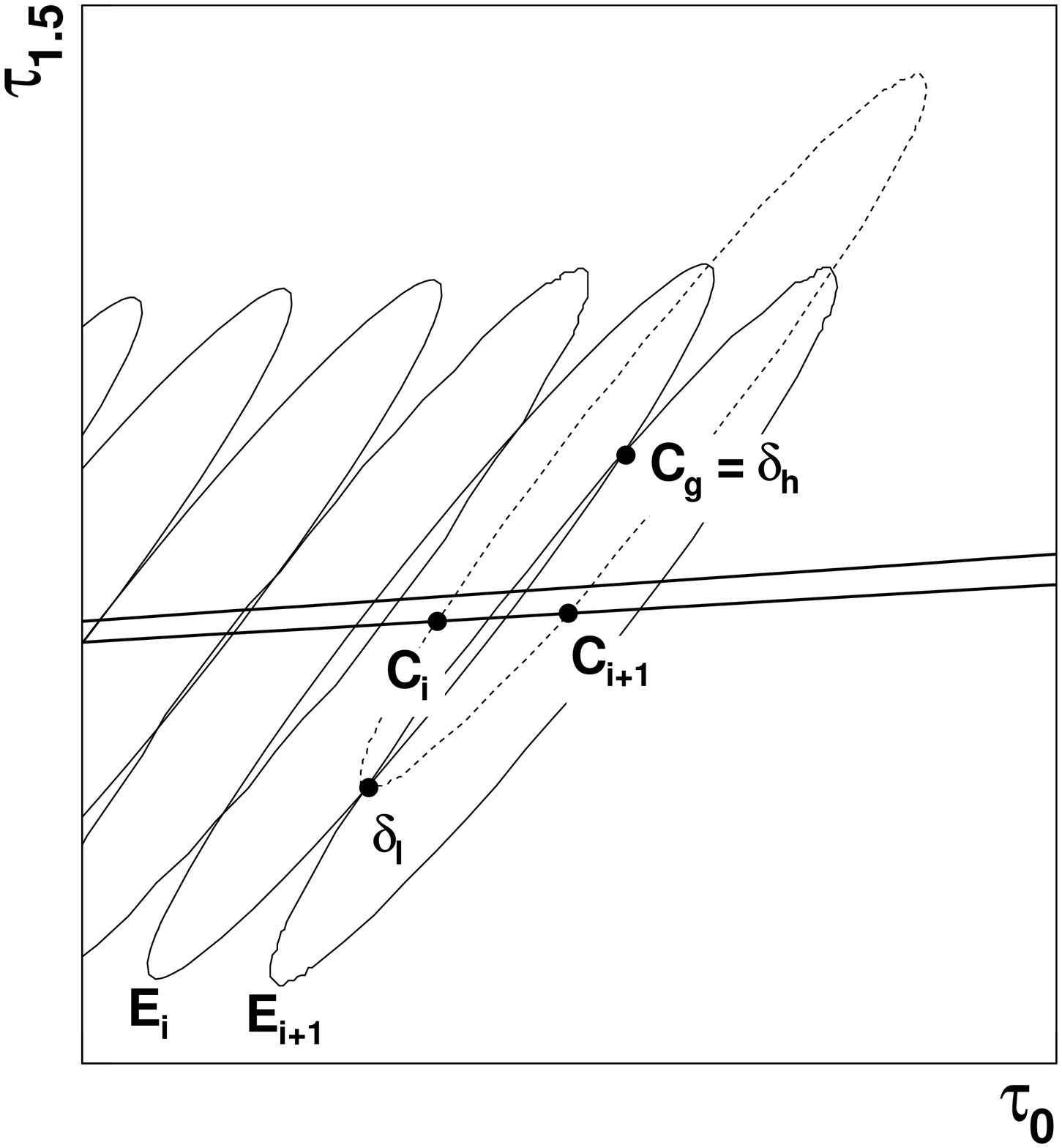}
} \\
\mbox{
\footnotesize{Choice of the position of the guiding ellipse}
} &
\mbox{
\footnotesize{Position of the next ellipse}
}
\end{tabular}
\end{center}
\caption{\label{placement_stage1_principle}\footnotesize{principle of the first stage
of placement along the equal mass line}}
\end{figure}

 The principle is described in figure \ref{placement_stage1_principle}. Starting from
an ellipse $E_i$ the center of which $C_i$ lies on the equal mass line, a choice
is made (explained hereafter) of the position $C_g$ of the center of a guiding ellipse along the
border of $E_i$. Because of the isomatch contour properties stated above,
$C_i$ lies on the guiding ellipse. It is also on the equal mass line. $C_{i+1}$
is the other intersection of the guiding ellipse and the equal mass line.

The position $C_g$ is chosen between $\lambda_1$ and $\lambda_2$ limits
on the $E_i$ ellipse, in such a way that the surface of the $\{C_i,C_{i+1},C_3\}$
triangle is maximized. $C_3$ is the location of the center of a potential ellipse $E_3$
that would form with $E_i$ and the next ellipse $E_{i+1}$ a three ellipse set
optimally placed (with the placement conditions imposed by the parameter
space lower boundary correponding to the equal mass line).
The $\lambda_1$ and $\lambda_2$ limits are chosen empirically and are subject
to the influence of numerical errors as well as interpolation/extrapolation
errors.
\par The next ellipse $E_{i+1}$ is then placed at position $C_{i+1}$.
$E_{i+1}$ and $E_i$ should ideally intersect at two points $\delta_h$
and $\delta_l$, $\delta_h$ being equal to the center of the guiding ellipse
$C_g$ and $\delta_l$ being in the physically forbidden region underneath the
equal mass line.
Because of the curvature and variation of the metric, it may happen that
$E_{i+1}$ and $E_i$ do not intersect. In that case, the position $C_{i+1}$
of $E_{i+1}$ is shifted towards $C_i$ along the equal mass line until the point
$C_g$ comes on $E_{i+1}$.
\par The first stage placement algorithm stops when $\delta_h$ falls inside
the parameter space, which means that two ellipses are needed to cover
the parameter space in the vertical direction.

   \subsection{Second stage}
\begin{figure}[htb]
\begin{center}
\includegraphics[height=50mm]{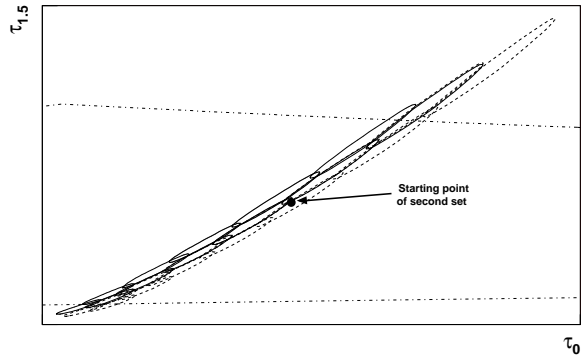}
\end{center}
\caption{\label{placement_three_lines}Independent placement of the third line
starting from a crossing point on the edge of a first two line set.}
\end{figure}

The second stage of placement consists of the coverage of the parameter space
line by line, as was described in \cite{buskulic_gwdaw02}.
\begin{itemize}
\item One starts from
a three ellipse set placed optimally at a point $D_0$.
\item Then place iteratively
ellipses using successive guiding ellipses that follow rules
defined in section \ref{isomatch_properties}. The placement is done alternatively
on the left and on the right of the line of guiding ellipses, and successively
above and below the initial point $D_0$.
\item One obtains a two-line set crossing
the parameter space (fig. \ref{placement_three_lines}).
Among the external crossings of the generated ellipses
of one of the lines (called $\gamma_{1i}$), a point $D_1$ is chosen and the
process is iterated.
\item At each step, only one of the lines is kept, the other being
approximately superimposed with a line generated at the previous step
(fig. \ref{placement_three_lines}).
\item The starting point of each two-line set $D_i$ for the step $j$ is chosen
among the $\gamma_{(j-1)i}$ as the point outside the parameter space and
not in the physically forbidden region which is the closest to the border
of the parameter space. Other choices have led to the observation of
variations in the direction of two successive lines, giving holes in the
coverage of the space.
\end{itemize}

\par The starting point for the first line building of the second stage is
the first intersection point $\delta_h$ found in the first stage that is
inside the parameter space. The placement ends when no ellipse from a line covers
any part of the parameter space.

   \subsection{Correcting points felt outside of the parameter space}
Once the first two stages are finished, a cleaning is performed to remove superfluous
ellipses that do not cover any part of the parameter space.
\par It is not possible to do it beforehand because it is not obvious if a given
ellipse covers or not a part of the parameter space before it is actually placed.
Its center may lie outside of the parameter space but a small part of the ellipse
may still cover a portion of the parameter space.
\par A position correction is also done on ellipses which, while covering a portion
of the parameter space, have their center in the physically forbidden region.
Those ellipses are shifted following the guiding contour used for their
generation until they fall on the equal mass border.

   \subsection{Examples of computed placements}
\begin{figure}[htb]
\begin{center}
\begin{tabular}{ccc}
\mbox{
\includegraphics[width=50mm, height=40mm]{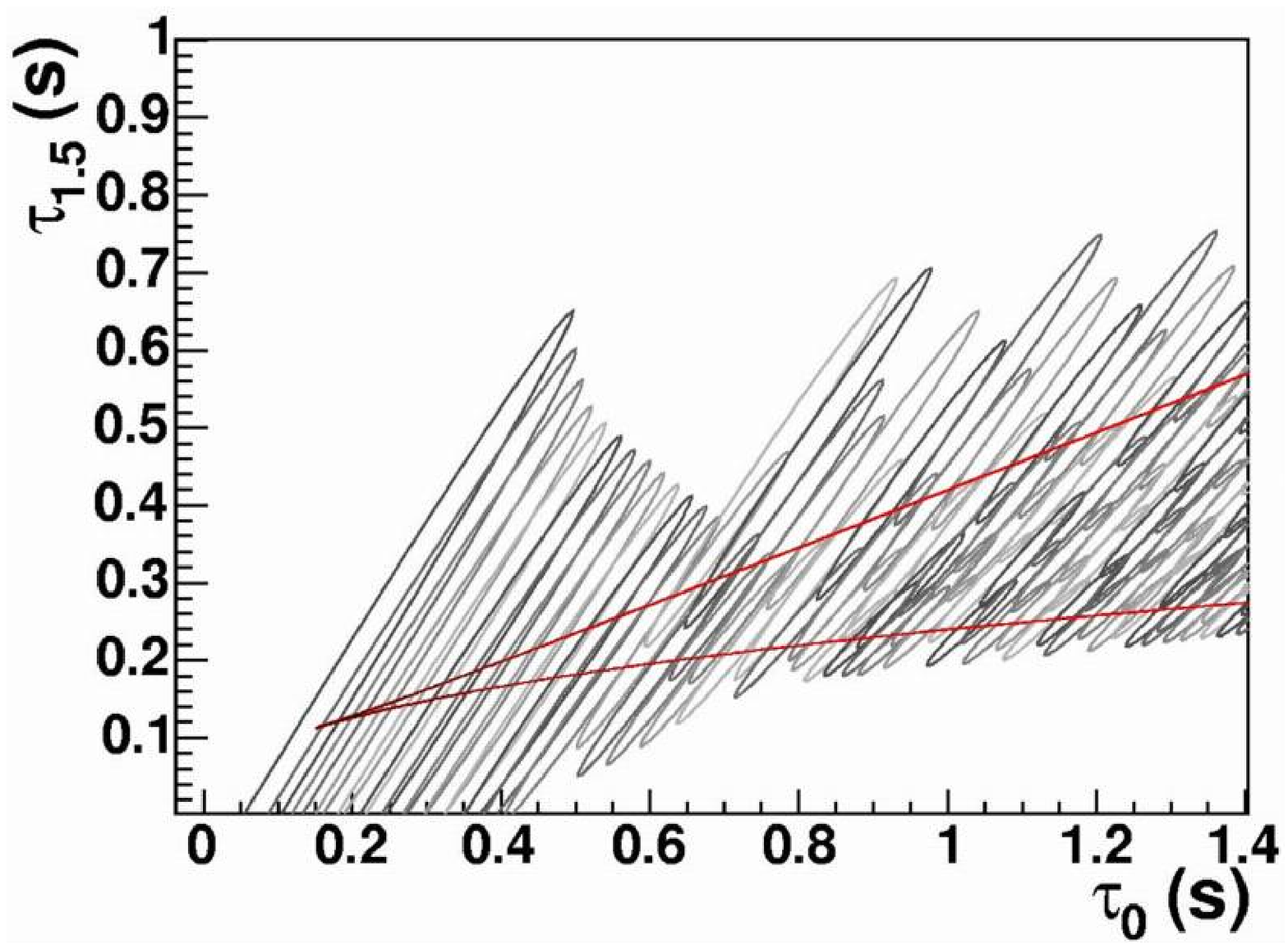}
} &
\mbox{
\includegraphics[width=50mm, height=40mm]{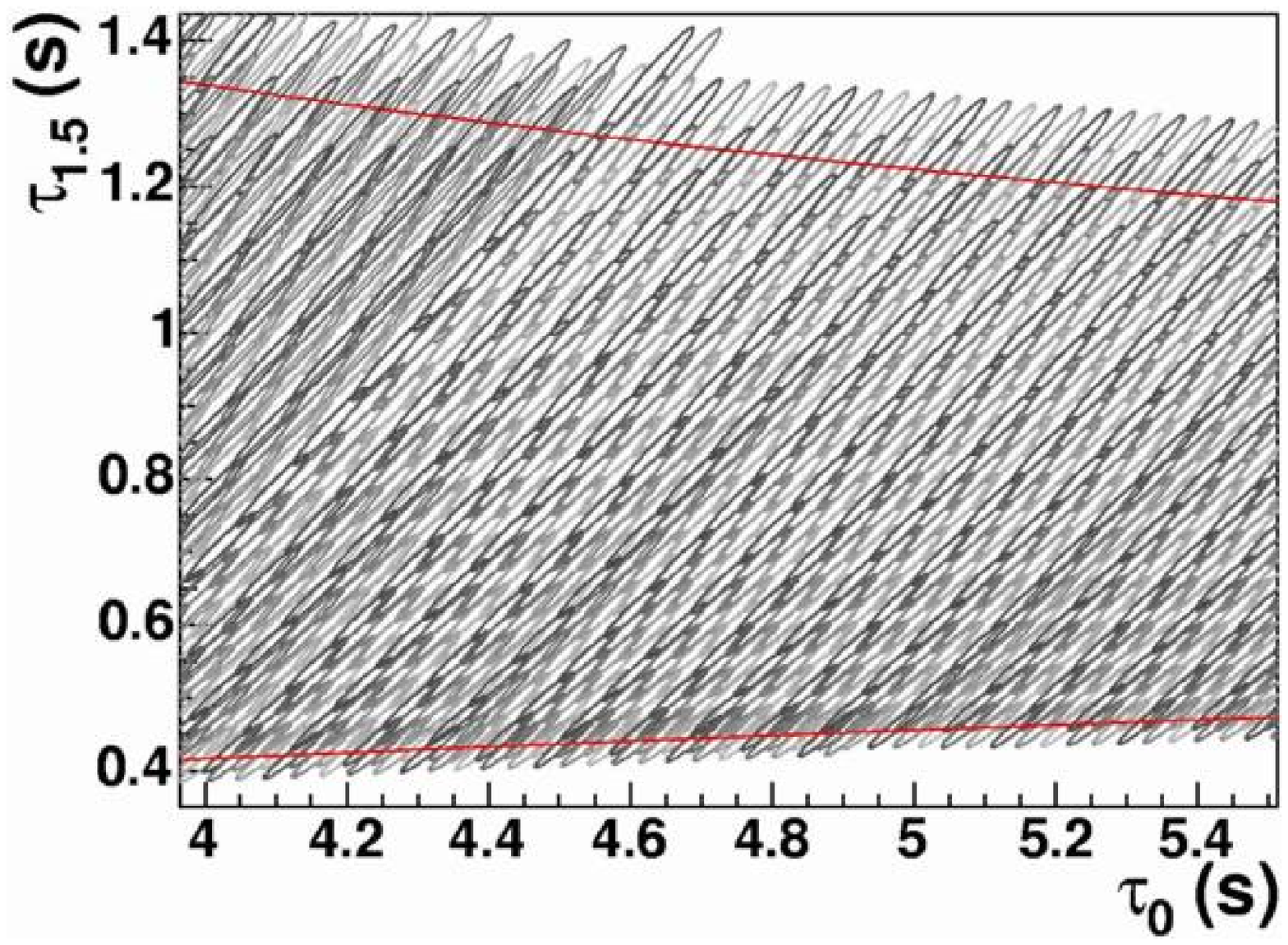}
} &
\mbox{
\includegraphics[width=50mm, height=40mm]{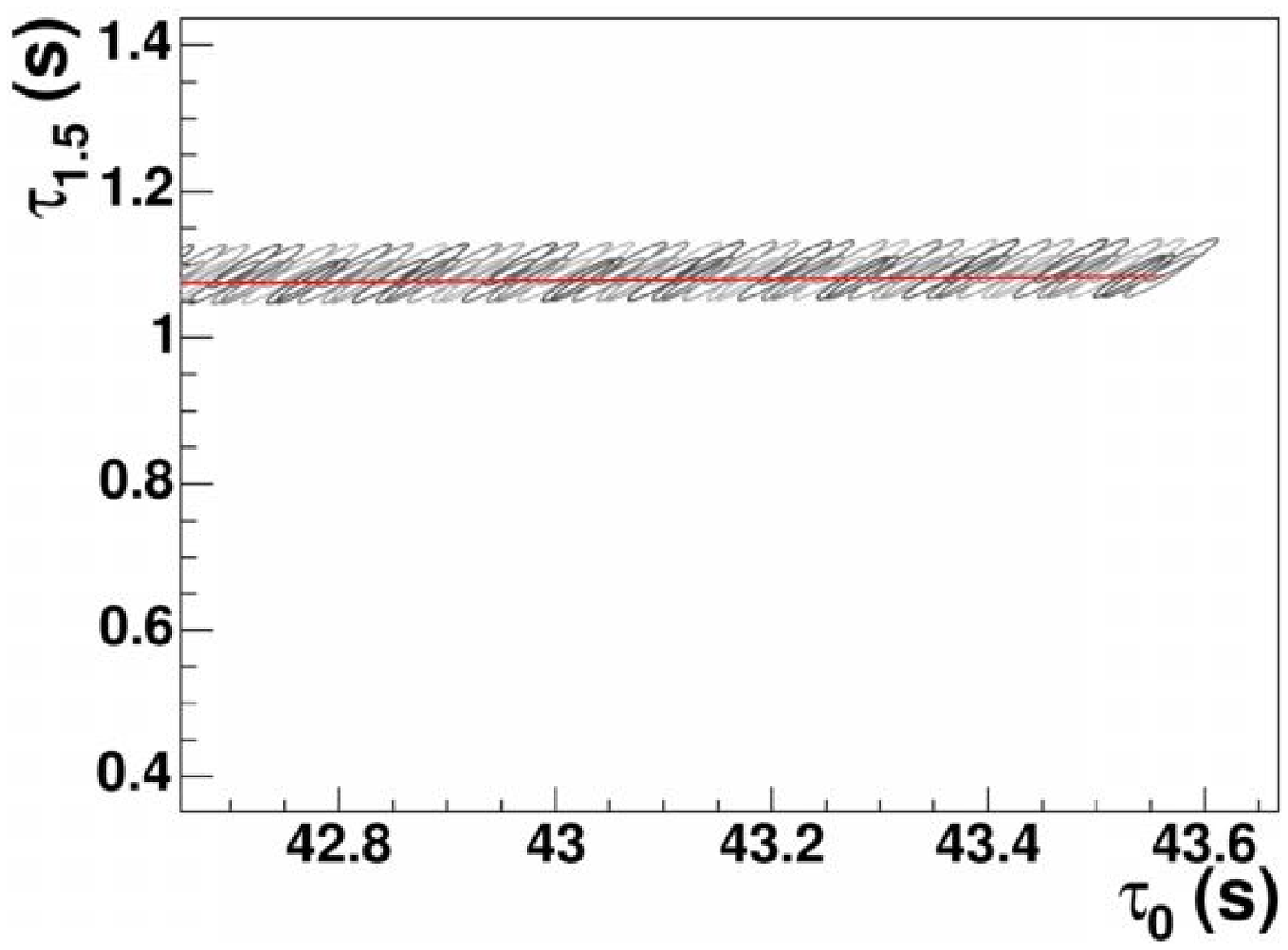}
}
\end{tabular}
\scriptsize{$m_{min}=1 \textrm{M}_\odot$, $m_{max}=30 \textrm{M}_\odot$, $MM=0.95$,
$F_l=40$ Hz, $F_h=2000$ Hz,$PN=2$, $N_T=152$, $N_P=11369$}
\end{center}
\caption{\label{placement_1_30}\footnotesize{Example of placement obtained for
real world CB searches in Virgo. The black line represents
the border of the parameter space. Three portions of space are shown
and color of ellipses is varied to help viewing the shapes.
See the text for an explanation of computing conditions.}}
\end{figure}

\begin{figure}[htb]
\begin{center}
\begin{tabular}{cc}
\mbox{
\includegraphics[width=70mm, height=45mm]{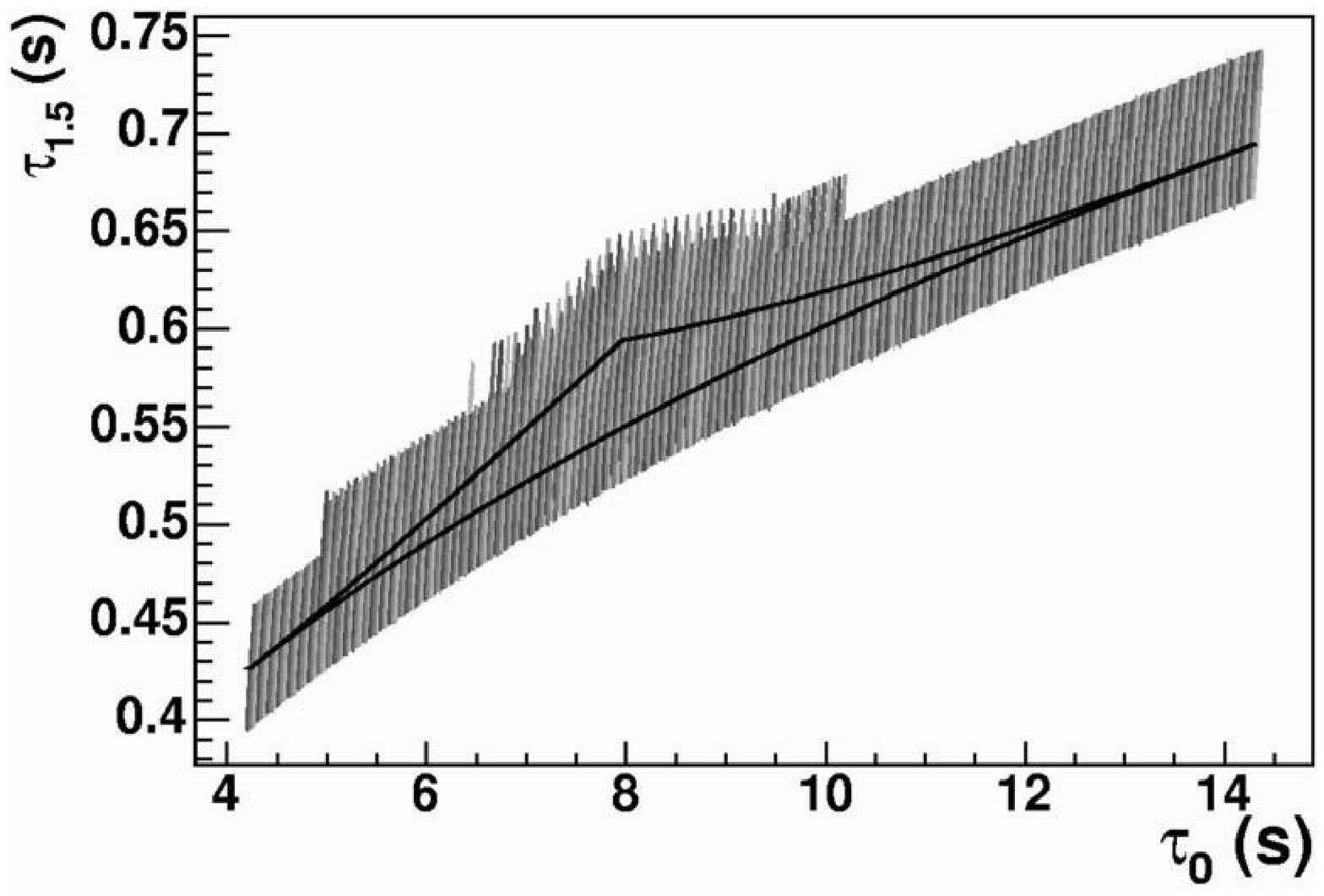}
} &
\mbox{
\includegraphics[width=70mm, height=45mm]{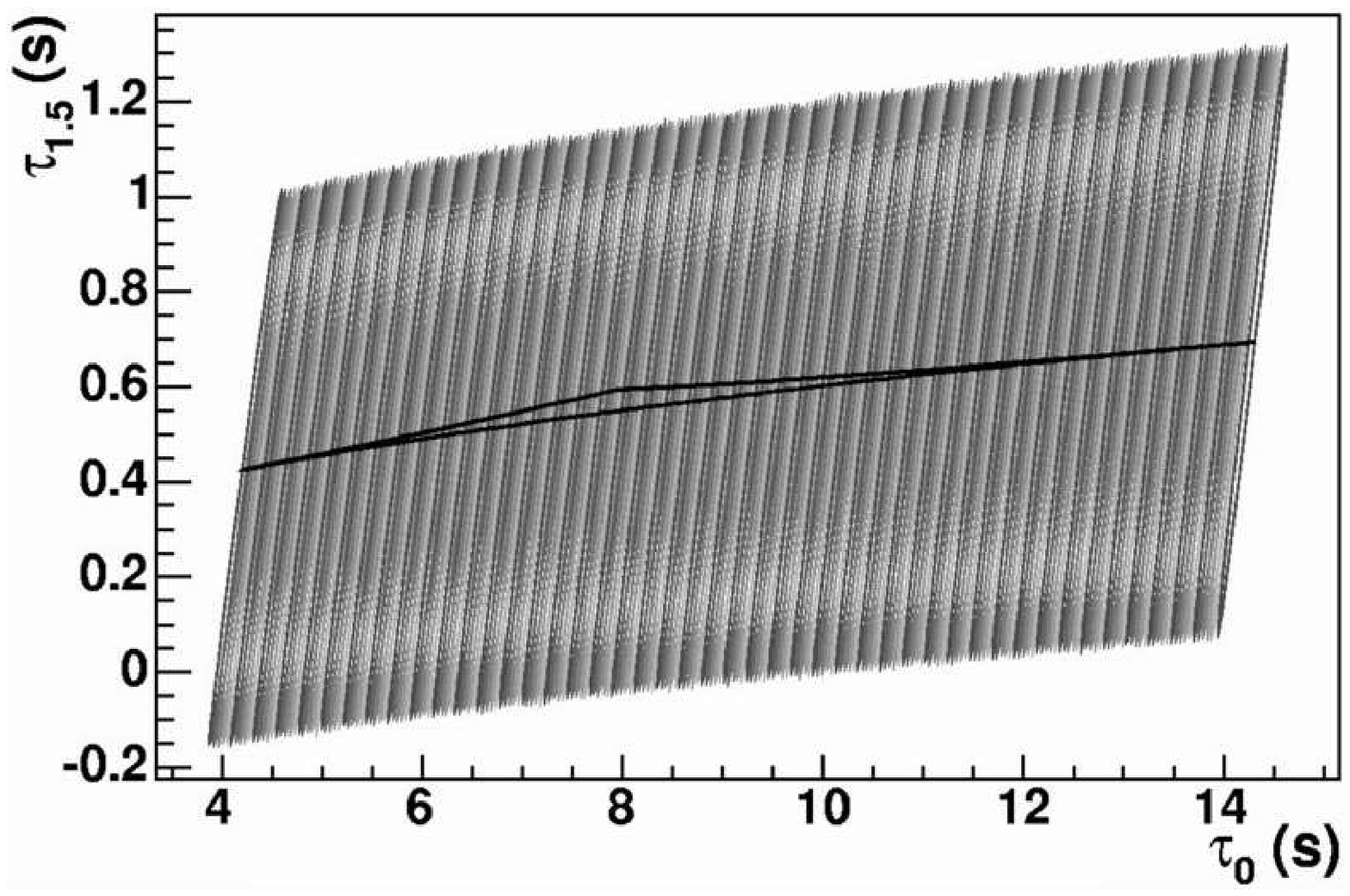}
} \\
\mbox{
\footnotesize{$F_l=40$ Hz, $F_h=2000$ Hz, $N_P=1011$}
} &
\mbox{
\footnotesize{$F_l=40$ Hz, $F_h=100$ Hz, $N_P=226$}
} \\
\mbox{
} &
\mbox{
\footnotesize{Only first stage needed to pave all space}
}
\end{tabular}
\newline
\newline
\footnotesize{$m_{min}=1.95 \textrm{M}_\odot$, $m_{max}=4.05 \textrm{M}_\odot$, $MM=0.95$,
$PN=2$}
\end{center}
\caption{\label{placement_2_4}\footnotesize{Example of placement obtained for
real world CB searches in Virgo. The black line represents
the border of the parameter space.
Color of ellipses is varied to help viewing the shapes.
See the text for an explanation of computing conditions.}}
\end{figure}

Figures \ref{placement_1_30} and \ref{placement_2_4} show a few real use-cases of placement.

\begin{itemize}
\item $m_{min}$ and $m_{max}$ are the minimal and maximal masses of the parameter
space
\item $MM$ the minimal match, $F_l$ and $F_h$ the lower and higher frequency
cutoffs used for the generation of templates
\item $PN$ is the order of the post-newtonian expansion
\item $N_T$ is the number of calculated points in the triangulation
\item $N_P$ is the number of points found in the placement.
\item The PSD used was a Virgo-like one (fig. \ref{spectrum_triang})
\end{itemize}

\section{Performance tests}
\label{sec_performance}

\subsection{Number of templates with the simple placement algorithm}
\label{nb_templates}
The number of templates needed for complete space coverage represents a simple
performance estimator. An estimation of this number was already given \cite{owen_2}
by computing the ratio between the volume of the parameter space and the proper volume
covered by a single template. It was supposed that the packing algorithm used was a square
(or hypercubic in $D$ dimensions) one. The proper volume is then, in 2 dimensions
\begin{equation}
\Delta V = 2(1-MM)
\end{equation}
and in the triangular lattice case (hexagonal Vorono\"i sets),
which was used in our simple algorithm for $D-2$
\begin{equation}
\Delta V = \frac{3\sqrt{3}}{2}(1-MM)
\end{equation}
Table \ref{n_templates_simple} shows the numbers we found by using an actual Virgo noise
power spectral density. $V_{ps}$ being the volume of the parameter space,
$N_P^{refSquare}$ is the reference number computed as in \cite{owen_2}
assuming a square packing, $N_P^{refTriang}$ assumes a triangular packing algorithm and
$N_P^{simple}$ is the actual number found with our simple algorithm
described in paragraph \ref{simple_pave}, which also produces
a triangular lattice. Edge effects appear clearly, as the smallest the volume of the parameter
space, the largest the difference between $N_P^{refTriang}$ and $N_P^{simple}$.

\begin{table}[hbt]
\begin{center}
\begin{tabular}{|c|||c|c|c|c|}
\hline
\mbox{\boldmath$m_{min}\textrm{ (M}_\odot\textrm{)}$} & \mbox{\boldmath$V_{ps}$} &
\mbox{\boldmath$N_P^{refSquare}$} & \mbox{\boldmath$N_P^{refTriang}$} &
\mbox{\boldmath$N_P^{simple}$}\\
\hline
\hline
0.5 & \mbox{$1.43\times10^4$} & \mbox{$1.43\times10^5$} & \mbox{$1.10\times10^5$} & \mbox{$1.28\times10^5$}\\
1 & \mbox{$2.57\times10^3$} & \mbox{$2.57\times10^4$} & \mbox{$1.98\times10^4$} & \mbox{$2.59\times10^4$} \\
3 & \mbox{$1.38\times10^2$} & \mbox{$1.38\times10^3$} & \mbox{$1.06\times10^3$} & \mbox{$1.80\times10^3$} \\
\hline
\end{tabular}
\caption{\label{n_templates_simple}\footnotesize{Comparison of the number of templates
obtained with a simple algorithm (triangular packing) and a theoretical number
(ratio between parameter space volume and proper volume asuming a square or triangular packing).
The minimum mass varies from
0.5 $\textrm{M}_\odot$ to 3 $\textrm{M}_\odot$. Other conditions~:
$m_{max}~=~30 \textrm{M}_\odot$,
$f_{min}~=~30 \textrm{Hz}$, $f_{max}~=~2000 \textrm{Hz}$, $MM~=~0.95$, Virgo-like PSD.}}
\end{center}
\end{table}

\subsection{Performance gain}
With the grid of templates coming out of the new placement algorithm,
one can expect a gain in the total computational cost needed to perform a search
over the defined parameter space with respect to the simple placement algorithm
(paragraph \ref{simple_pave}). This gain is not easily quantified because
it depends on the specific search algorithm and on aspects that do not depend on the
computational algorithm itself, such as I/O. But it may be estimated in at least two ways:
\begin{itemize}
\item firstly, by the gain in the overall number of templates coming out of the placement
algorithms (method A).
\item secondly, by modeling the ``standard'' method for doing the optimal
filtering and searching for an approximation of the gain (method B).
\end{itemize}
\par The optimal
filtering technique and an estimation of the computational cost are described by Schutz
in \cite{blair}. An approximation of the cost (number of floating point operations)
for analyzing a set of $N_{tot}$ data values for a given template of length $N_s$,
and a fractional overlap $x$ of successive data set chunks is:

\begin{equation}
N_{flop} = \frac{N_{tot}}{1-x}[3ln_2\frac{N_f}{x}+4]
\end{equation}

A discussion on the optimal value of $x$ is made in \cite{blair}, but it
does not take into account the I/O costs, as well as exchanges of data
between memory and processor, which is found to be critical in our case.
Therefore, as explained in \cite{vicere}, we choose $x$ so as to roughly
optimize the length and the number of the vectors to be exchanged between
the core memory and the CPU.
Starting from the expression above and fixing $x=\frac{1}{2}$ for each template,
it may be shown that the approximate total number of operations needed to analyze a set
of $M$ templates is given by:

\begin{equation}
N_{flop}^T = 6N_{tot}\sum_{i=1}^{M}[ln_2(2f_s\tau_i)+\frac{4}{3}]
\end{equation}

where $f_s$ is the sampling frequency and $\tau_i$ the length of an individual template.
This leads to consider a computing performance estimator of the form

\begin{equation}
\label{equXi}
\xi = \sum_{i=1}^{M}[ln_2(2f_s\tau_i)+\frac{4}{3}]
\end{equation}

Since we only want an approximate expression, we consider
$\tau_i=\tau_0(i)$, where $\tau_0(i)$ is the newtonian chirp time
of the coalescing binary producing a given template.
We made comparisons between the placement produced by the simple method
of paragraph \ref{simple_pave} and the full placement method. Tables \ref{ngain_mm},
\ref{ngain_mmin} and \ref{ngain_f}
show the gain on the number of templates and the gain on the
performance estimator $\xi$. The conditions of the tests were varied but the base
conditions were the following:
\label{standard_conditions}
\begin{itemize}
\item minimal mass $m_{min}=1 \textrm{M}_\odot$,
\item maximal mass $m_{max}=30 \textrm{M}_\odot$
\item minimal frequency for template generation $f_{min}=30 \textrm{Hz}$
\item maximal frequency for template generation $f_{max}=2000 \textrm{Hz}$
\item minimal match $MM=0.95$
\item power spectral density close to the final Virgo one (fig. \ref{spectrum_triang})
\end{itemize}
\par In the tables, $N_P^{refSquare}$ is the reference number of templates, computed
as in \cite{owen_2} assuming a square packing, as explained above in section
\ref{nb_templates}. $N_P$ represents the number of templates found by the placement
to cover the parameter space, $G_{N_P}=(N_P^{simple}-N_P^{full})/N_P^{simple}$
is the gain in the number of templates obtained when going from the simple placement
to the full placement method, $N_T$ is the
number of seed templates necessary to triangulate the parameter space,
$G_{\xi}$ is the gain in performance estimator. Unless otherwise noted,
the triangulation process was stopped after 7 steps of refinement,
which was shown in the section \ref{global_view} not to bring problems.
\begin{table}[hbt]
\begin{center}
\begin{tabular}{|c|||c|c|c|c|||c|c|c|c|}
\hline
\mbox{\footnotesize \boldmath$MM$} & \mbox{\footnotesize \boldmath$N_P^{refSquare}$} &
\mbox{\footnotesize \boldmath$N_P^{simple}$} & \mbox{\footnotesize \boldmath$N_P^{full}$} & \mbox{\footnotesize \boldmath$G_{N_P}$} &
\mbox{\footnotesize \boldmath$N_T$} & \mbox{\footnotesize \boldmath$\xi_{simple}$} & \mbox{\footnotesize \boldmath$\xi_{full}$} &
\mbox{\footnotesize \boldmath$G_{\xi}$}\\
\hline
\hline
\footnotesize 0.90 & \footnotesize \it{12840} & \footnotesize 14047 & \footnotesize \bf{10746} & \mbox{\footnotesize \boldmath$23.5\%$} & \footnotesize 395
   & \mbox{\footnotesize $2.46\times10^5$} & \mbox{\footnotesize \boldmath$1.91\times10^5$} & \mbox{\footnotesize \boldmath$22.4\%$} \\
\footnotesize 0.95 & \footnotesize \it{25680} & \footnotesize 26183 & \footnotesize \bf{20161} & \mbox{\footnotesize \boldmath$23.0\%$} & \footnotesize 381
   & \footnotesize \mbox{$4.58\times10^5$} & \mbox{\footnotesize \boldmath$3.57\times10^5$} & \mbox{\footnotesize \boldmath$21.9\%$} \\
\footnotesize 0.98 & \footnotesize \it{64200} & \footnotesize 61144 & \footnotesize \bf{47183} & \mbox{\footnotesize \boldmath$22.8\%$} & \footnotesize 394
   & \mbox{\footnotesize $1.06\times10^6$} & \mbox{\footnotesize \boldmath$8.34\times10^5$} & \mbox{\footnotesize \boldmath$21.6\%$} \\
\hline
\end{tabular}
\caption{\label{ngain_mm}\footnotesize{Gain on the number of templates and on performance
coefficient with respect to a simple algorithm and varying the minimal match.
Other conditions: $m_{min}~=~1 \textrm{M}_\odot$, $m_{max}~=~30 \textrm{M}_\odot$,
$f_{min}~=~30 \textrm{Hz}$, $f_{max}~=~2000 \textrm{Hz}$, Virgo-like PSD.}}
\end{center}
\end{table}

\par In table \ref{ngain_mm}, the minimal match was varied from $MM=0.90$ to $MM=0.98$,
keeping the other parameters equal. As can be seen, an average performance gain of roughly
22\% is achieved. It may be noted that the number of templates may also be used as
a performance estimator, giving numbers very similar to $\xi$.

\begin{table}[hbt]
\begin{center}
\begin{tabular}{|c|||c|c|c|||c|c|c|c|}
\hline
\mbox{\boldmath$m_{min}\textrm{ (M}_\odot\textrm{)}$} &
\mbox{\boldmath$N_P^{simple}$} & \mbox{\boldmath$N_P^{full}$} & \mbox{\boldmath$G_{N_P}$} &
\mbox{\boldmath$N_T$} & \mbox{\boldmath$\xi_{simple}$} & \mbox{\boldmath$\xi_{full}$} &
\mbox{\boldmath$G_{\xi}$}\\
\hline
\hline
0.5 & 129507 & \bf{113531} & \mbox{\boldmath$12.3\%$} & 335
   & \mbox{$1.28\times10^6$} & \mbox{\boldmath$1.13\times10^6$} & \mbox{\boldmath$11.9\%$} \\
1 & 26183 & \bf{20161} & \mbox{\boldmath$23.0\%$} & 381
   & \mbox{$4.58\times10^5$} & \mbox{\boldmath$3.57\times10^5$} & \mbox{\boldmath$21.9\%$} \\
3 & 1829 & \bf{1106} & \mbox{\boldmath$39.5\%$} & 416
   & \mbox{$1.65\times10^4$} & \mbox{\boldmath$1.01\times10^4$} & \mbox{\boldmath$38.7\%$} \\
\hline
\end{tabular}
\caption{\label{ngain_mmin}\footnotesize{Gain on the number of templates and on performance
coefficient with respect to a simple algorithm. The minimum mass varies from
0.5 $\textrm{M}_\odot$ to 3 $\textrm{M}_\odot$. Other conditions:
$m_{max}~=~30 \textrm{M}_\odot$,
$f_{min}~=~30 \textrm{Hz}$, $f_{max}~=~2000 \textrm{Hz}$, $MM~=~0.95$, Virgo-like PSD.}}
\end{center}
\end{table}

\par In table \ref{ngain_mmin}, only the minimal mass of the stars, hence the size of
the parameter space, was varied from $m_{min}=0.5\textrm{M}_\odot$ to
$m_{min}=3\textrm{M}_\odot$. The gain is naively expected to increase with the size
of the parameter space. The bigger the parameter space, the higher the variation of metric,
hence the bigger the variation in size of the ellipses. The results shown in table
\ref{ngain_mmin} vary in the opposite direction. This is explained by edge effects, where
the influence of ellipses covering a small part of the parameter space, on or outside the
border, and the way they are placed, play a dominant role.

\begin{table}[hbt]
\begin{center}
\begin{tabular}{|c|||c|c|c|||c|c|c|c|}
\hline
\mbox{\boldmath$f_{min}-f_{max}\textrm{ (Hz)}$} &
\mbox{\boldmath$N_P^{simple}$} & \mbox{\boldmath$N_P^{full}$} & \mbox{\boldmath$G_{N_P}$} &
\mbox{\boldmath$N_T$} & \mbox{\boldmath$\xi_{simple}$} & \mbox{\boldmath$\xi_{full}$} &
\mbox{\boldmath$G_{\xi}$}\\
\hline
\hline
30-100 & 2191 & \bf{1948} & \mbox{\boldmath$11.1\%$} & 90
   & \mbox{$4.09\times10^4$} & \mbox{\boldmath$1.37\times10^4$} & \mbox{\boldmath$11.2\%$} \\
100-2000 & 986 & \bf{833} & \mbox{\boldmath$15.5\%$} & 167
   & \mbox{$1.37\times10^3$} & \mbox{\boldmath$1.17\times10^3$} & \mbox{\boldmath$15.2\%$} \\
30-2000 & 12641 & \bf{11820} & \mbox{\boldmath$6.6\%$} & 61
   & \mbox{$2.33\times10^5$} & \mbox{\boldmath$2.18\times10^5$} & \mbox{\boldmath$6.5\%$} \\
\hline
\end{tabular}
\caption{\label{ngain_f}\footnotesize{Gain on the number of templates and on performance
coefficient with respect to a simple algorithm. Two limited frequency ranges are tested:
$[30;100]$ Hz and $[100;2000]$ Hz. The mass range is
$[1;5]\ \textrm{M}_\odot$. Other conditions: 
$MM~=~0.95$, Virgo-like PSD.}}
\end{center}
\end{table}

Finally, table \ref{ngain_f} shows the results for a variation in the frequency
range. The mass range was limited to $[1;5]\ \textrm{M}_\odot$ because for high masses we are
reaching the limits of the numerical relevance of the metric calculation.

\par It may be noted that in practical algorithms, templates will be
grouped by groups of similar length. The expression of $\xi$ (equation \ref{equXi})
will take a linear form as a function of the number of templates. This should
bring the gains we obtained for the performance estimator closer to the ones obtained
with the number of templates.

\subsection{Coverage tests}
The metric calculation is approximate, especially in the high mass region,
where there is yet no good model of coalescence. It is therefore
important to do independent tests on the covering efficiency. Monte-Carlo tests were
performed by testing randomly scattered points over the parameter space. The distribution
of position is uniform in $(\tau_0,\tau_{1.5})$ parameters. For each point, the corresponding
waveform is computed and the match with the templates of the bank is calculated,
retaining the highest. Actually, only the subset of templates which are closer
than a given distance to the point, in the metric sense, are considered.
\par
\begin{figure}[htb]
\begin{center}
\begin{tabular}{cc}
\mbox{
\includegraphics[height=50mm]{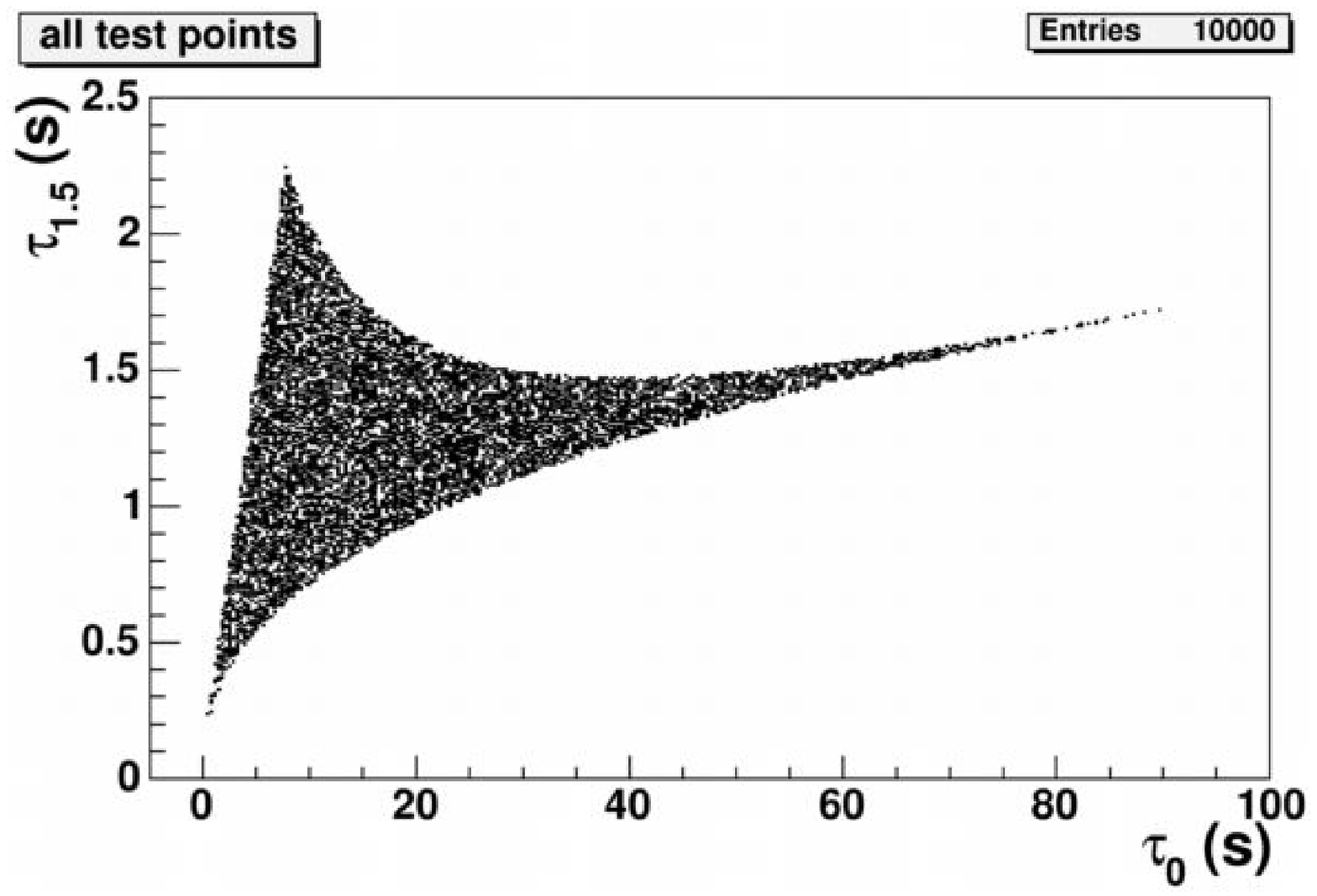}
} &
\mbox{
\includegraphics[height=50mm]{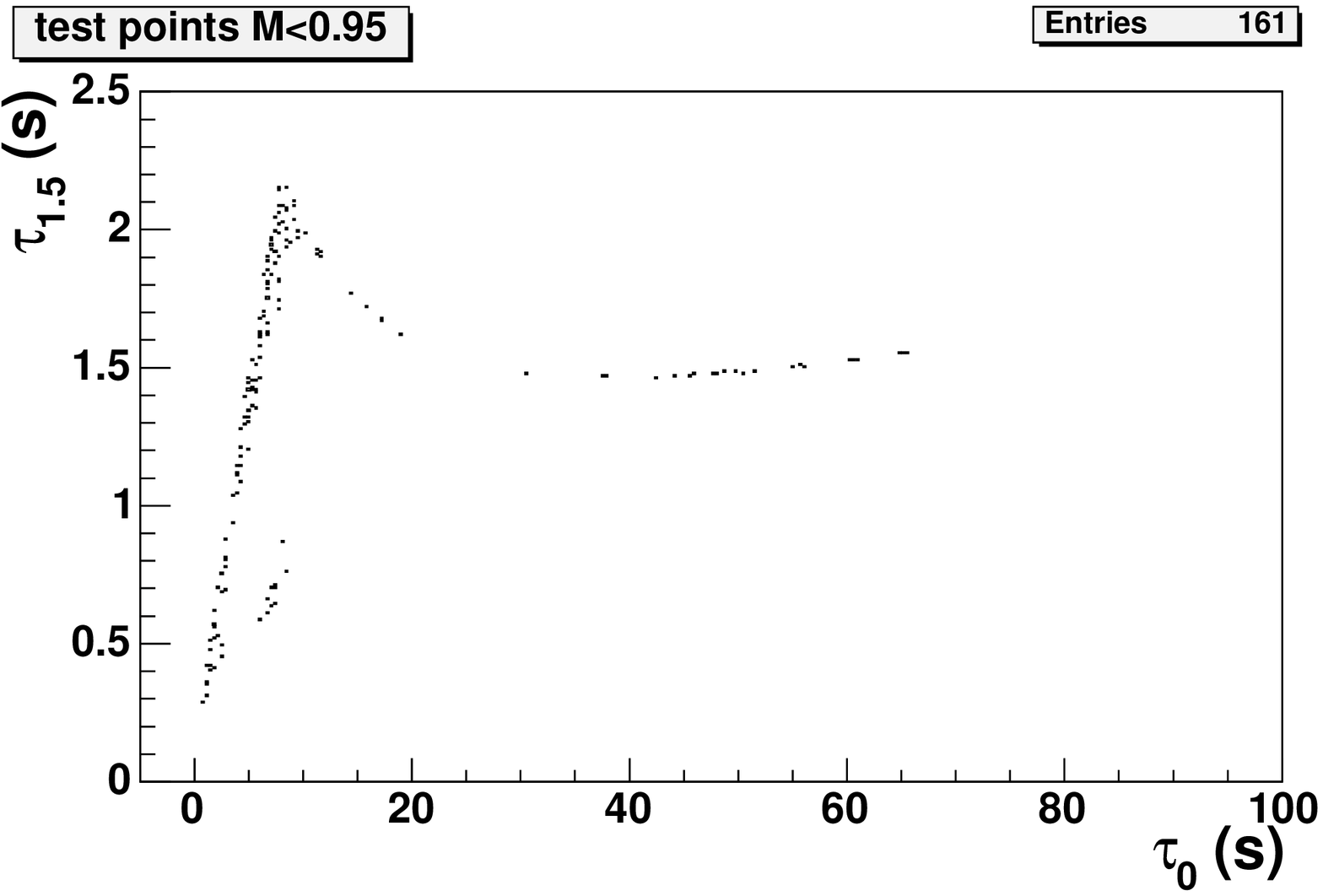}
} \\
\mbox{
\footnotesize{A}
} &
\mbox{
\footnotesize{B}
} 
\end{tabular}
\end{center}
\caption{\label{matchtestpoints}\footnotesize{Monte-Carlo distribution of points in 
coverage tests. A: all the points, B: points with match $M<0.95$ with the closest
template. The conditions are the standard ones.}}
\end{figure}

 The chosen conditions in terms of masses, frequencies and minimal match
are the standard ones described in \ref{standard_conditions}.
Figure \ref{matchtestpoints}.A shows the distribution of test points
over the parameter space, while figure \ref{matchtestpoints}.B shows
the distribution of points the match of which is lower than the specified match
(0.95 in our case).

\begin{figure}[htb]
\begin{center}
\includegraphics[width=90mm]{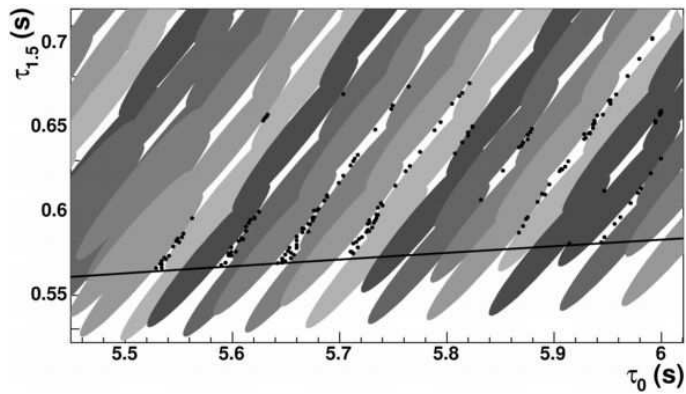}
\end{center}
\caption{\label{testinhole}\footnotesize{Superposition of a small part of
calculated isomatch ellipses in placement and test points with match $M<0.95$.
The test points clearly fall in holes formed by locally incorrect placement.}}
\end{figure}

The low match points (with match $M<0.95$) represent 1.6\% of all the test points. There
are two possible reasons for the presence of these points. The first is the presence of holes
in between ellipses, due to suboptimal placement,
the second is a possible miscalculation of the metric in some peculiar cases,
for example for high mass binaries.
Finer Monte-Carlo tests were performed in small regions relevant for the two cases,
and low match point positions were superimposed with isomatch ellipses.
The first case is illustrated with figure \ref{testinhole} where it is clearly
seen that most of the low match points fall in existing holes of the placement.

\begin{figure}[htb]
   \begin{minipage}[t]{0.45\linewidth}
      \begin{center}
      \includegraphics[height=50mm, width=75mm]{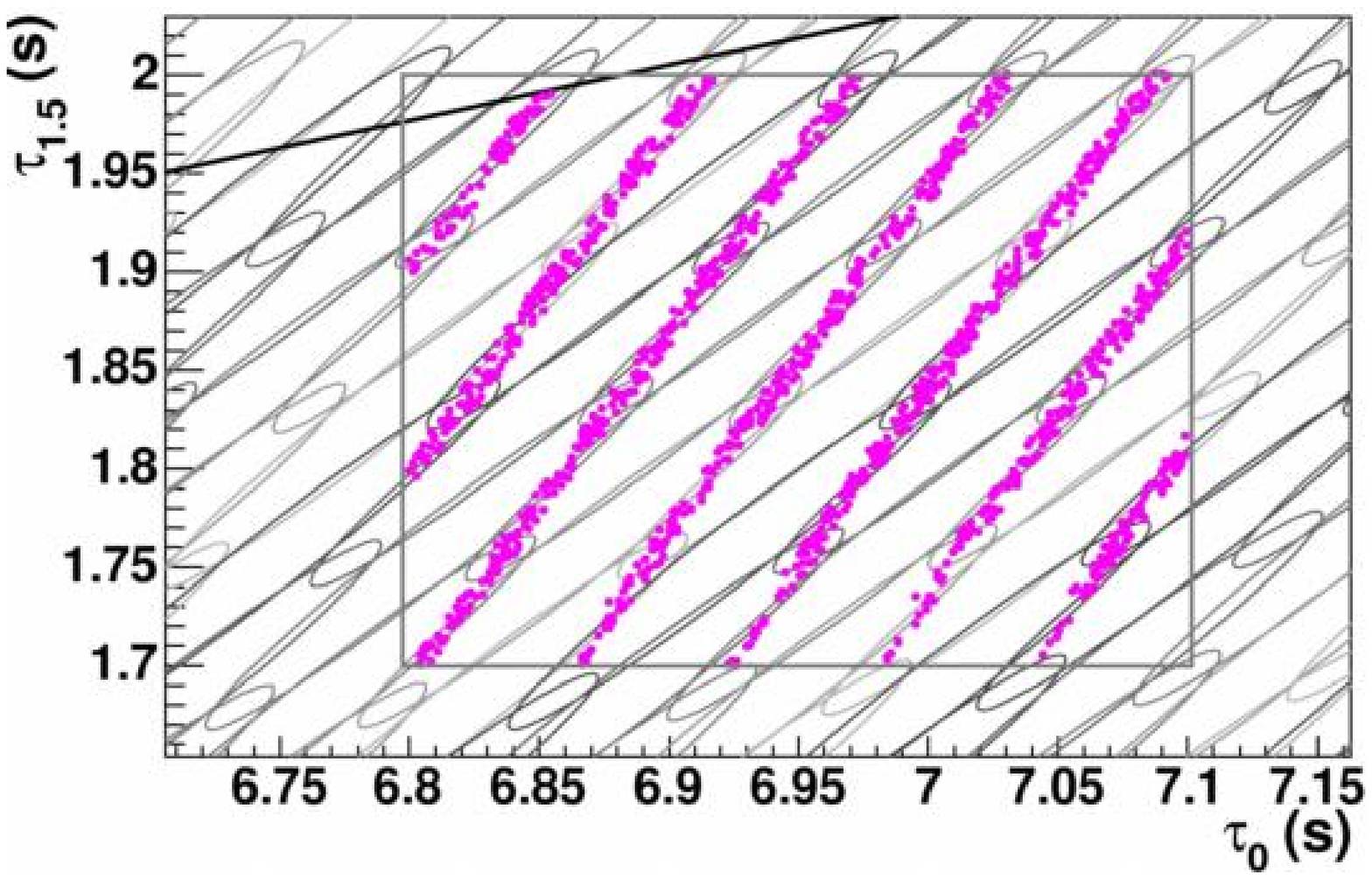}
      \end{center}
      \caption{\label{testhighmass_out}\footnotesize{Superposition of a small part of
calculated isomatch ellipses in placement and test points with match $M<0.95$. The
gray box surrounds the test region.}}
   \end{minipage}
   \hspace{0.1\linewidth}
   \begin{minipage}[t]{0.45\linewidth}
      \begin{center}
      \includegraphics[height=50mm, width=75mm]{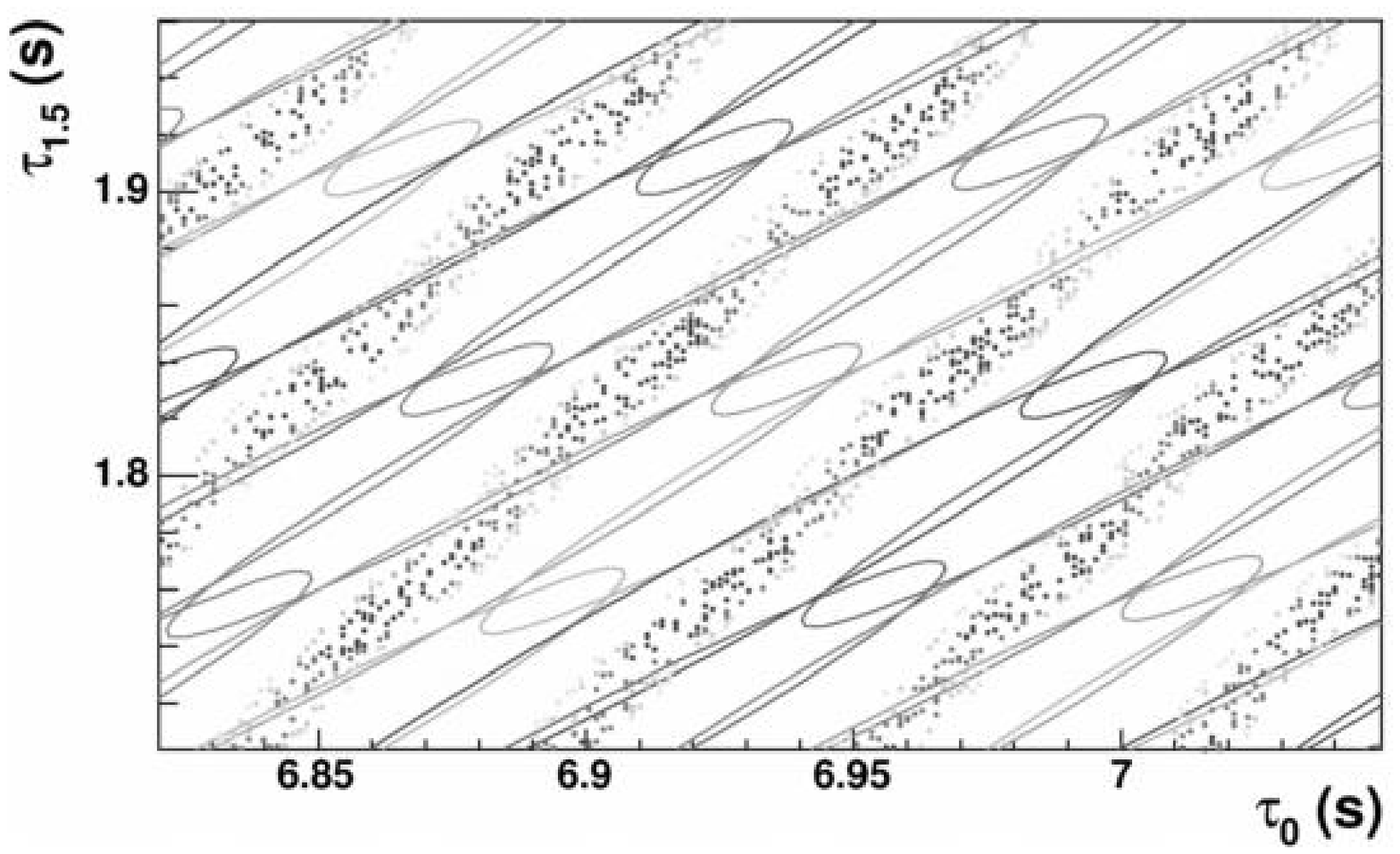}
      \end{center}
      \caption{\label{testhighmass_in}\footnotesize{Superposition of a small part of
calculated isomatch ellipses in placement and test points with match $M>0.99$.
This illustrates the wrongly calculated orientation of the ellipses, leading to a wrong
placement.}}
   \end{minipage}
\end{figure}

The second case is illustrated in figure \ref{testhighmass_out}. The test
is made with points chosen in the region $\tau_0 \in [6.8;7.1]$ and
$\tau_{1.5} \in [1.7;2]$ (region named $\sigma_{HM}$), the points being
inside the parameter space.
It is clear from the picture that all the points of $\sigma_{HM}$ fall well inside
an existing ellipse, hence they should have had match $M>0.95$ if
the metric was correctly calculated.
This situation is explained by the miscalculation of the ellipse orientation,
as is illustrated in figure \ref{testhighmass_in}. In this figure, the points with
$M>0.99$ were considered, and they form a figure clearly showing the wrong
orientation of the computed ellipses (several colors depending on the value
of the match were used, the darker the points the higher the match).

\par
\begin{figure}[htb]
   \begin{minipage}[t]{0.45\linewidth}
      \begin{center}
      \includegraphics[height=50mm]{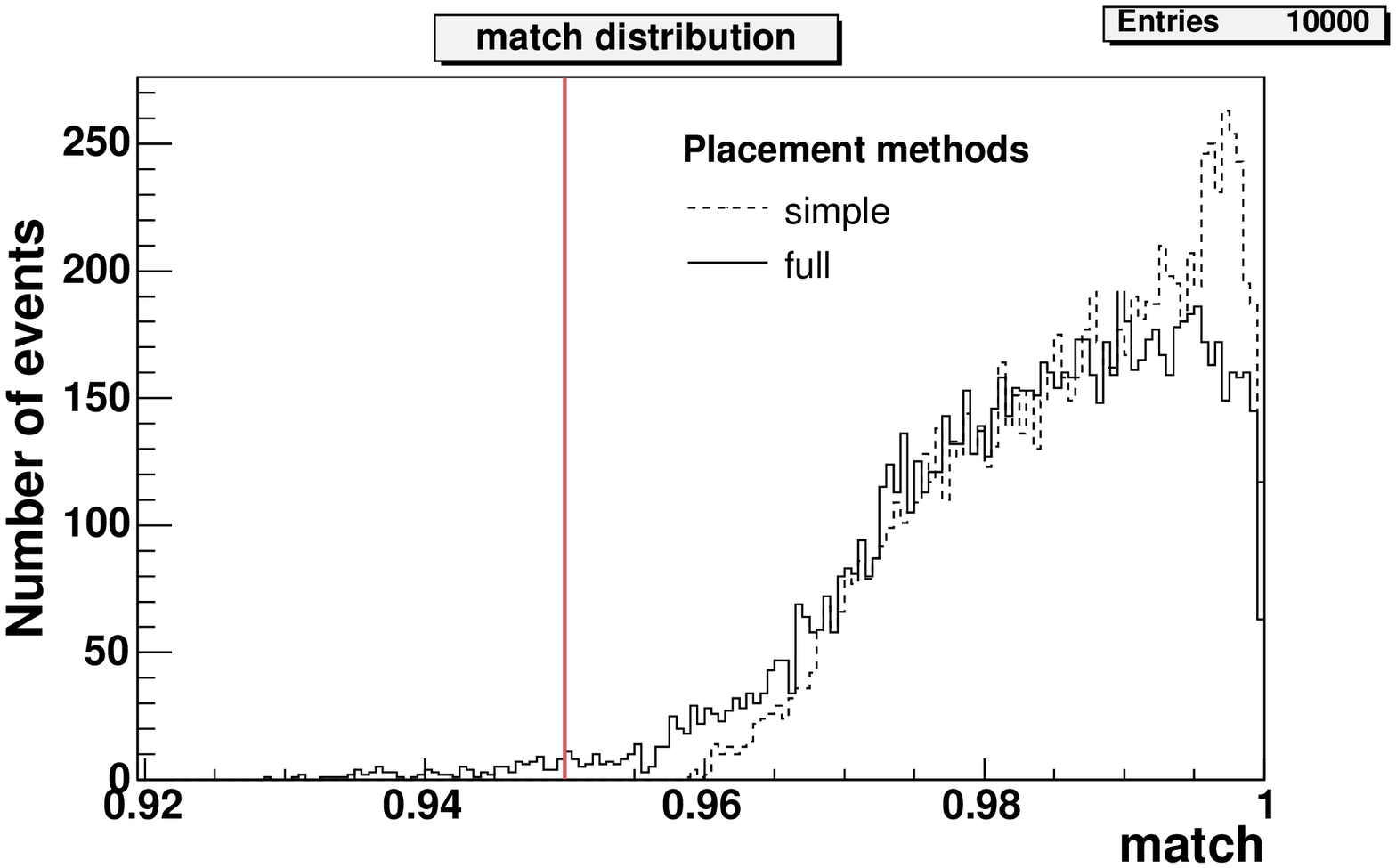}
      \end{center}
      \caption{\label{testdistrib}\footnotesize{Comparison of the match
distribution for the simple and full placement. The red vertical line
shows the requested minimal match, 0.95 in this case.}}
   \end{minipage}
   \hspace{0.1\linewidth}
   \begin{minipage}[t]{0.45\linewidth}
      \begin{center}
      \includegraphics[height=50mm]{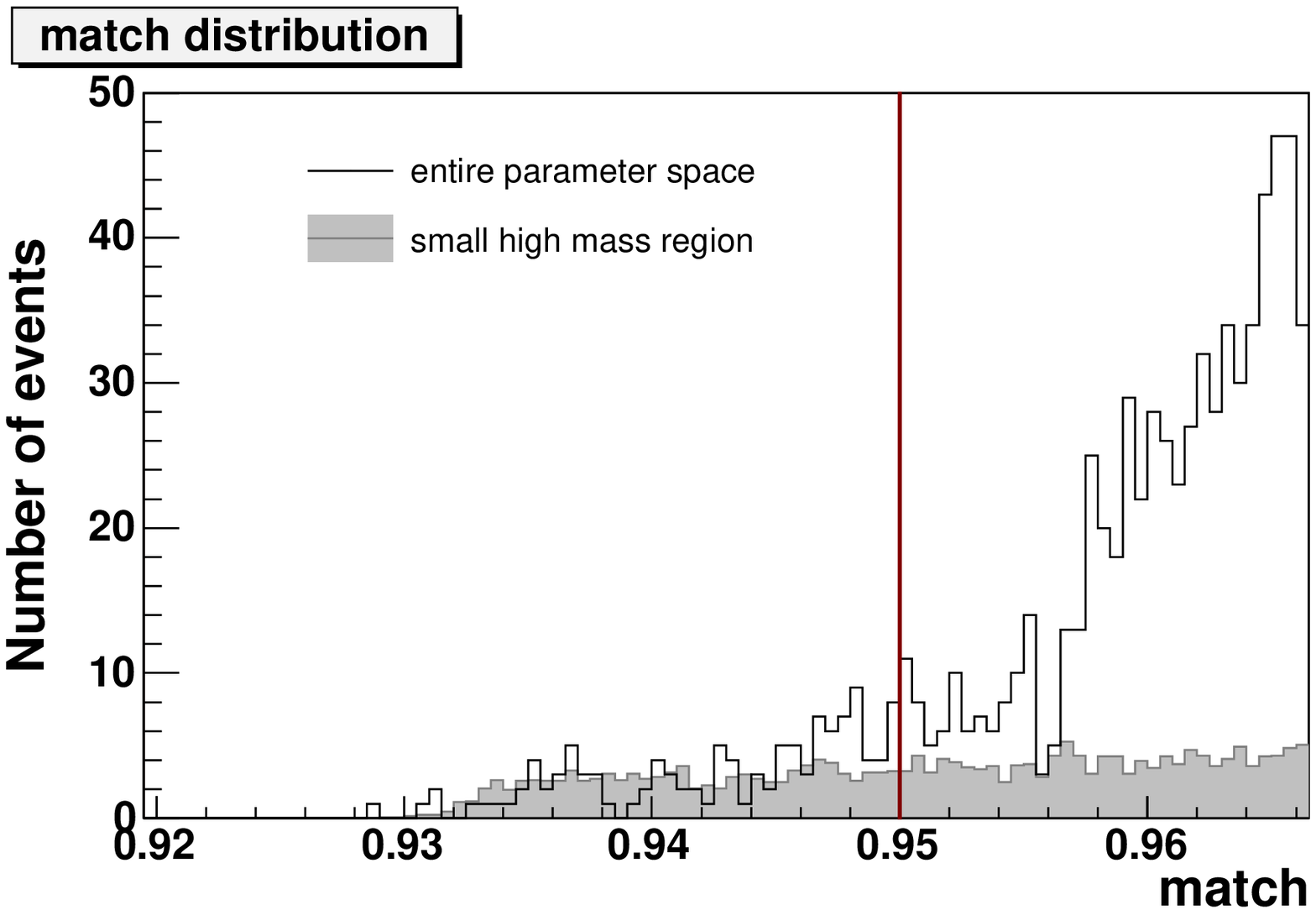}
      \end{center}
      \caption{\label{testdistrib_limited}\footnotesize{Estimation of the
contribution of points in a region where ellipses correspond to high mass
with miscalculated metric. The red vertical line
shows the requested minimal match, 0.95 in this case.}}
   \end{minipage}
\end{figure}

 Figure \ref{testdistrib}
compares the distribution of the test points match
for the full placement algorithm and for the simple
placement algorithm. The simple placement is clearly suboptimal, but
ensures a complete covering of the parameter space while the optimality
is better for the full algorithm, though it does not cover perfectly
the parameter space, at the level of a few percent undercoverage.
Figure \ref{testdistrib_limited} illustrates the influence of miscalculation
of the metric. Superimposed to the distribution of the match in the
full placement case, is the distribution of the match for $\sigma_{HM}$.
This distribution was scaled down
proportionately to the surface of $\sigma_{HM}$ region versus
the surface of the parameter space to show
its contribution to the overall distribution.

\par In general, the two effects, miscalculation and misplacement, are both present
with various strength throughout the whole parameter space. Miscalculation
is due to wrong approximation of the metric and/or approximations in
the triangulation and interpolation steps of the placement algorithm.
\par
\begin{figure}[htb]
\begin{center}
\includegraphics[width=110mm]{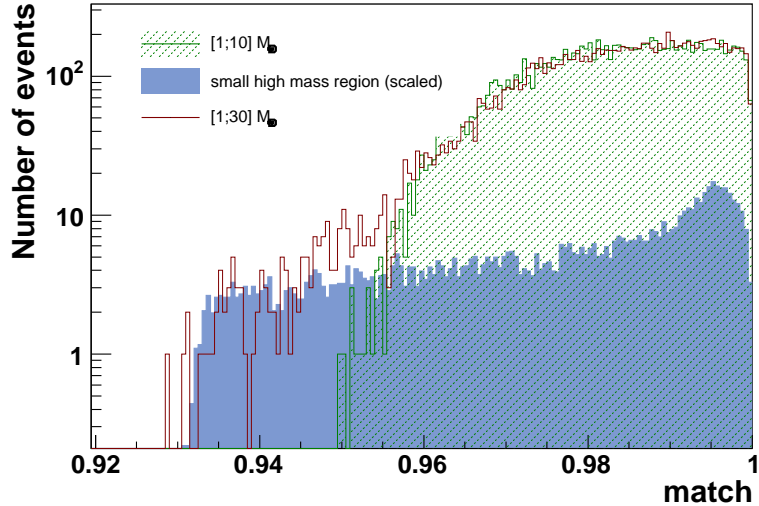}
\end{center}
\caption{\label{testdistrib_all}\footnotesize{Comparison of match distributions
for $[m_{min};m_{max}]=[1;30] \textrm{M}_\odot$ and
$[m_{min};m_{max}]=[1;10] \textrm{M}_\odot$. Also included is the distribution
corresponding to a specifically high mass region in the parameter space, scaled
proportionately to its surface with respect to that of the whole parameter space.}}
\end{figure}

 In figure \ref{matchtestpoints}.B, a clear accumulation of low match points
seems to occur in the high mass region. To confirm this, a test was made with
a mass range $[m_{min};m_{max}]=[1;10] \textrm{M}_\odot$. The match distribution
for this test, superimposed on the match distribution of the entire parameter space
($[m_{min};m_{max}]=[1;30] \textrm{M}_\odot$) and on the match distribution of
$\sigma_{HM}$, is shown in figure \ref{testdistrib_all}.
It is clear from this figure that the main source of low match points
is the miscalculation of the metric for high masses.

\par
\begin{figure}[htb]
\begin{center}
\includegraphics[width=110mm]{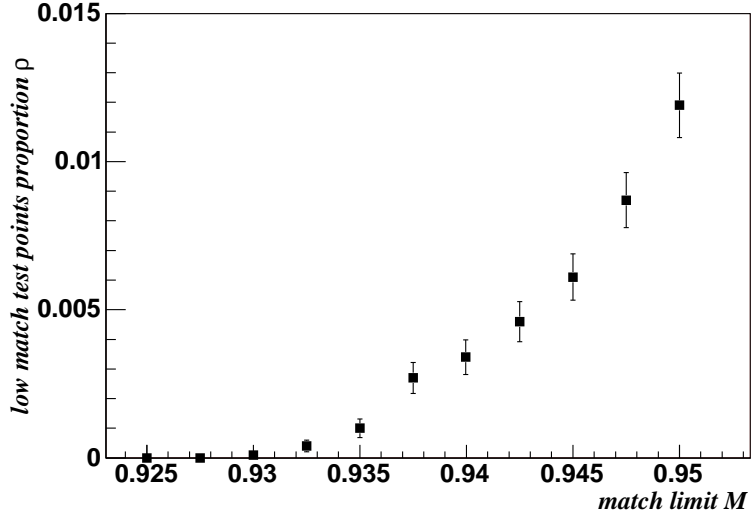}
\end{center}
\caption{\label{efficiency_full}\footnotesize{Proportion of bad match test points as a
function of an effective minimal match value for a given placement.
Only statistical errors are reported.}}
\end{figure}

 In order to get an idea of how to easily overcome these problems, one can
calculate the proportion of bad match test points as a function of
a varying minimal match value $\mathpzc{M}$, for a given placement. This corresponds
to enlarging the ellipses obtained with a placement with an initial
minimal match $\mathpzc{M}_0$. 
Figure \ref{efficiency_full} shows the variation of
the proportion $\rho$ of test points with match lower than $\mathpzc{M}$
versus $\mathpzc{M}$. From this figure, given a desired bad match points proportion,
one gets an estimation of the effective minimal match reached.

A question may be raised about the robustness of the algorithm, i.e. is
the algorithm adequate for real, noisy data. It is very difficult to assess
an ``absolute'' robustness of the algorithm, because of three main points :

\begin{itemize}
\item the difference between the true contour and the calculated ellipse,
especially for low matches, which may in some circumstances push the
algorithm to its limits.
\item the fact that the algorithm is not robust in the case of large and
fast variations of the metric.
\item the difference between calculated and interpolated ellipses that
may, though not in large proportions, affect the algorithm.
\end{itemize}

There is a need for tools that run online and verify the relevance
of the computed grid banks. In the case of problems, it is always possible
to switch to the simple algorithm.

\subsection{Speed tests and recomputation of the placement}

All the tests were performed on a Linux 2.4 GHz Pentium IV workstation
and we present in table \ref{compute_time} the computation time in seconds
needed for each placement, in increasing number of generated grid points.
The time is divided in two, corresponding to the two main steps of the
algorithm, namely first the triangulation and generation of seed contours
and second the placement itself. There is a rough proportionality between
the number of final grid points and the time, with a quasi constant
term corresponding to the first step (the number of generated seed contours
being always of the same order of magnitude).
\begin{table}[hbt]
\begin{center}
\begin{tabular}{|c|c|c|c|c|}
\hline
\mbox{\boldmath$m_{min}-m_{max}\textrm{ (M}_\odot\textrm{)}$} & \mbox{\boldmath$MM$} &
\mbox{\boldmath$N_T$} & \mbox{\boldmath$N_P^{full}$} & Time (s)\\
\hline
\hline
3-30 & 0.90 & 418 & \bf{643} & 60 \\
3-30 & 0.95 & 416 & \bf{1106} & 67 \\
3-30 & 0.98 & 402 & \bf{2362} & 84 \\
1-30 & 0.90 & 395 & \bf{10746} & 203 \\
1-30 & 0.95 & 381 & \bf{20161} & 294 \\
1-30 & 0.98 & 394 & \bf{47183} & 575 \\
0.5-30 & 0.90 & 335 & \bf{59103} & 742 \\
0.5-30 & 0.95 & 346 & \bf{113531} & 1287 \\
\hline
\end{tabular}
\caption{\label{compute_time}\footnotesize{Computation time for different placements on a
2.4 GHz Pentium IV Linux workstation.}}
\end{center}
\end{table}

\par The frequency of recomputation of the placement is still under consideration in Virgo.
It depends on the change rate of the shape of the sensitivity curve over time, the stability
of which is not yet fully assessed for future science runs. The numbers given in
table \ref{compute_time} may seem too large for a frequent recomputation, for instance
every 15 minutes, in the case of large volume parameter spaces.
Though such a frequency is not expected for the final Virgo science runs, we may
need to consider a parallelization of the algorithm. The part of the algorithm that
could be parallelized efficiently is the placement part, but one should not expect
more than an estimated factor 2 to 5 improvement in overall computing time, due to the
sequential nature of the algorithm. Indeed, in one line of ellipses, ellipse number $n$
may not be placed before ellipse number $n-1$. Only the placement of complete lines may be
somewhat decorrelated.

\section{Comparison with previous studies and perspectives}
Beside very important pioneering efforts \cite{owen}\cite{owen_2} the results of which
are now widely used, several previous studies were done for the template placement problem.
We believe that our method is somewhat complementary to them. For example, the placement
algorithm used in \cite{IUCAA_Pune} for extended hierarchical searches is based on a square
tiling. This is justified in this case by the low minimal match value used ($\Gamma = 0.8$),
which gives very irregularly shaped contours. Our method could probably be adapted
to such a case by applying methods such as in \cite{buskulic_gwdaw02} to determine
the shape of the contours, but an important effort has to be made to improve the
speed of the shape reconstruction algorithm, which is going to be one of the main limiting
factors.
\par Another example is the paper of Arnaud et al.\cite{arnaud} where authors
devise a 2D tiling method and
test it in the case of supernova ringdown signals. It is very difficult to make
a direct comparison between this algorithm and ours. The very large parameter space
curvature described by Arnaud et al is likely to bring some holes if we apply directly
our tiling method to ringdown signals. This would imply the need for an improvement
to our placement procedure. On the other hand, the Arnaud 2D tiling method
was not yet applied to the case of a $(\tau_0,\tau_{1.5})$
inspiral parameter space and it is not clear what would be the
result in terms of speed and possible overcoverage.
\par The computational geometry tools that we used are still valid in higher
dimensional spaces. It may be tempting to consider the extension of our
algorithm to multidimensional searches. In that case, the main challenge would be
to improve the algorithm speed, since the number of contours in nD
is roughly going as
\begin{equation}
N_n \approx N_2^{\frac{n}{2}}
\end{equation}
Where $N_2$ is the number of contours obtained in 2D. This is of course a ``worst case''
scenario where the granularity is the same (and high) in all dimensions. 

\section{Conclusion}
We presented a technique for doing the placement of isomatch
ellipses on a template parameter space using triangulation and
interpolation of seed ellipses. A comparison is done with a simple
regular triangular tiling using a single ellipse.
This comparison shows an improvement between 6\% and 30\% depending
on the mass range and frequency range. Some coverage tests were
also performed that show a few percent undercoverage of the parameter space,
mainly in the high mass region. This undercoverage seems to come
from the miscalculation of the metric for high masses. Finally,
speed tests were made.

\section{Acknowledgements}
We would like to thank all our Virgo colleagues from the inspiral
data analysis group, and particularly Andrea Vicer\'e for his help
in bringing an important piece of Mathematica code and
insightful comments. We would also like to acknowledge the use of
the Ligo Analysis Library, and thank Thomas Cokelaer for his precious
help.


\section*{Appendix: A few notions of computational geometry}
Since computational geometry is not very commonly used in our field, we will give
a very short introduction to the notions useful for the present study. It is
in no way exhaustive or pretending to be accurate. More details
may be found in \cite{rourke} or \cite{george}.

   \subsection*{Definition of a triangulation}
Given a set $S$ of points in a euclidian space, 2-dimensional in our case, we would like
to subdivide the space into a set of triangles, each triangle being formed by three points
from $S$. Any point $P$ in the space belongs to (is included into)
one and only one triangle. This is however not enough and the properties of
the set of triangles should be the ones of a {\em triangulation}.
\par
\begin{figure}[htb]
\begin{center}
\includegraphics[width=130mm]{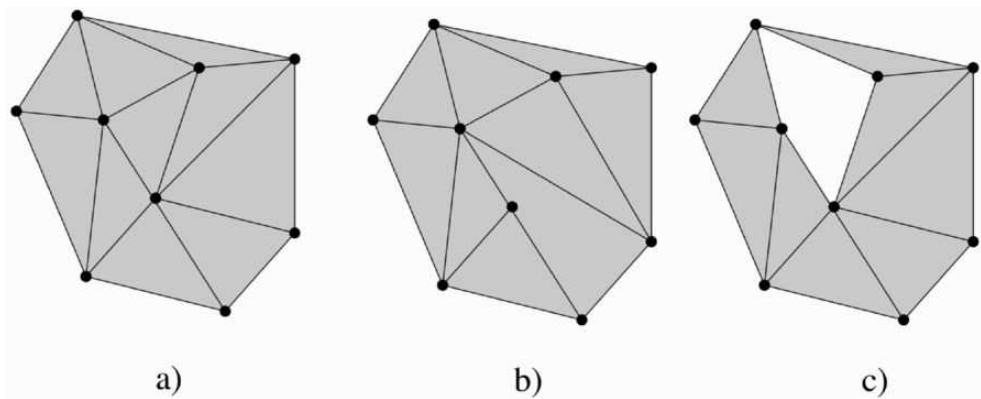}
\end{center}
\caption{\label{example_triang}\footnotesize{a) a valid triangulation, b) invalid since there are
two triangles sharing only a part of a face, c) invalid because part of the domain
is not covered by triangles}}
\end{figure}

 Some definitions first. Let's consider a set of $(n+1)$ affinely independent
points in an n-dimensional euclidian space $\mathbb{R}^n$.
\begin{itemize}
\item The convex hull of a set of points is the minimal convex set
containing all the points (imagine a rubber band stretched so that it
encompasses all the points).
\item A {\em simplex} is the convex hull of a set of $n+1$ points
(a line segment in 1D, a triangle in 2D, a tetrahedron in 3D,...).
\end{itemize}
A triangulation $T$ of the set of points $S$ in $\mathbb{R}^n$ is a subdivision of
$\mathbb{R}^n$ into n-dimensional simplices such that:
\begin{itemize}
\item The set of points that are vertices of the simplices coincides with $S$.
\item Any two simplices in $T$ intersect in a common face, only one vertex or not at all.
\item The convex hull of $S$ defines a domain $\Omega$ in $\mathbb{R}^n$.
If $K$ is a simplex, then
\begin{equation} \label{triangle_cover}
\Omega = \bigcup_{K \in T} K
\end{equation}
\end{itemize}
We illustrate the above definition by showing what is and what is not a triangulation
in a 2-dimensional space in figure \ref{example_triang} for a given set of points.

   \subsection*{Vorono\"{\i} diagram}
\begin{figure}[htb]
\begin{center}
\includegraphics[height=45mm]{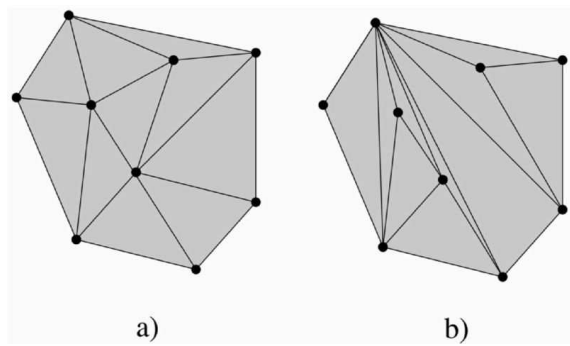}
\end{center}
\caption{\label{two_triangulations}\footnotesize{Two examples of valid triangulations
for the same set of points. Intuitively, the a) case is ``better'' than the b) case}}
\end{figure}

\begin{figure}[htb]
\begin{center}
\includegraphics[height=45mm]{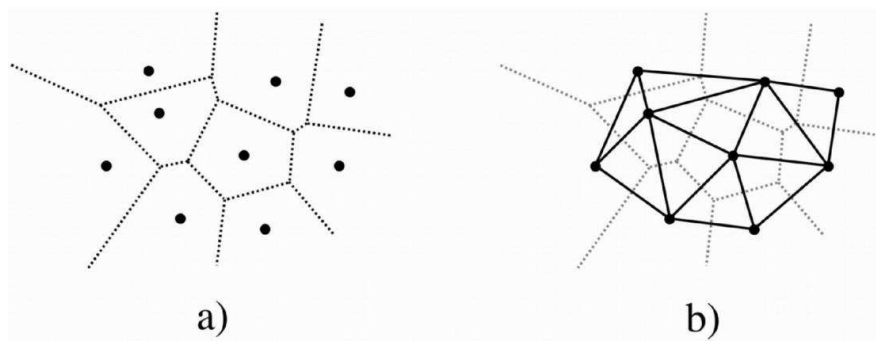}
\end{center}
\caption{\label{voronoi_diagram}\footnotesize{a) the Vorono\"{\i} diagram of a set of points
and b) the dual of the Vorono\"{\i} diagram, the Delaunay triangulation}}
\end{figure}

A triangulation is not unique, as may be seen in figure \ref{two_triangulations}.
All triangulations are not equivalent for a given problem. There is a need to define
a criterion of suitability. The most commonly used criterion is the Delaunay criterion
which constraints the compactness of the triangles and will be explained later.
It is linked to the so called
Vorono\"{\i} diagram. Given $\mathcal{S}$ a set of points $P_i$ in a $d$-dimensional space,
the Vorono\"{\i} diagram is the set of cells $\mathcal{V}_i$ associated with each point $P_i$
and defined as
\begin{equation} \label{voronoi_cell_def}
\mathcal{V}_i=\{P\in \mathbb{R}^n \textrm{  such that  } d(P,P_i)\leq d(P,P_j), \forall j\neq i\}
\end{equation}
Where $d$ is the euclidian distance between two points.
In other words, $\mathcal{V}_i$ is the locus of points in $\mathbb{R}^n$ closer to
$P_i$ than to any other point of $\mathcal{S}$.
It has been shown \cite{delaunay} that the geometrical dual of the Vorono\"{\i} diagram
is a triangulation, the Delaunay triangulation (fig. \ref{voronoi_diagram}).

   \subsection*{Delaunay triangulation}
\begin{figure}[htb]
\begin{center}
\includegraphics[height=45mm]{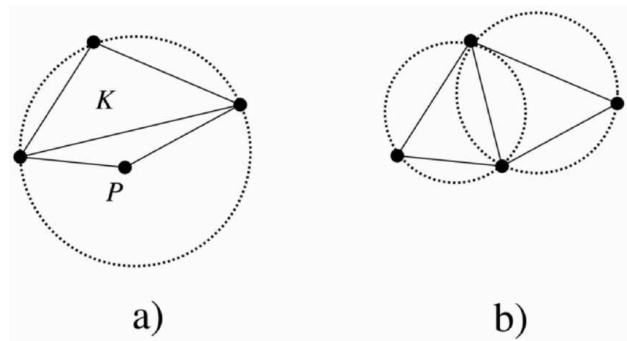}
\end{center}
\caption{\label{delaunay_example}\footnotesize{Example of a) a triangulation not
satisfying the Delaunay criterion (the point $P$ is inside
the circumdisk of $K$) b) satisfying it. A few circumcircles
of triangles are shown}}
\end{figure}

The Delaunay criterion states that the open circumdisk
(in 2 dimensions, circumsphere in $n$ dimensions) of a triangle (simplex) contains
no point from the set.
The example in figure \ref{delaunay_example} shows a triangulation not satisfying
the Delaunay criterion.
Among all possible triangulations, the Delaunay triangulation
\begin{itemize}
\item maximizes the minimum angle formed by the faces of the triangles
\item minimizes the biggest diameter of the circumcircles associated with the triangles
\end{itemize}

Intuitively, this would mean that the Delaunay triangulation produces the more ``compact''
triangles.
 
   \subsection*{A simple algorithm}
Based on the previous definition of the Delaunay criterion, it is possible to devise
a simple algorithm to compute a triangulation based on a set of points. It is called
an incremental algorithm, or Bowyer-Watson algorithm \cite{watson}\cite{bowyer}.
\par
\begin{figure}[htb]
\begin{center}
\includegraphics[width=80mm]{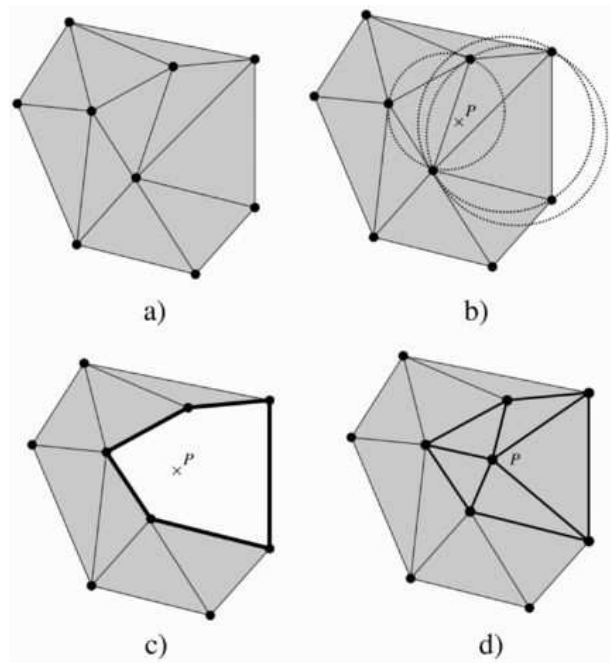}
\end{center}
\caption{\label{incremental_algo}\footnotesize{Incremental algorithm. a) add a point $P$ to an existing
triangulation, b) remove all triangles whose circumcircle contains $P$, c) obtain a polygon
enclosing P, d) triangulate only this polygon}}
\end{figure}

 The algorithm is incremental in the sense that the points of the set $S$ are added
one by one, recomputing a triangulation at each step.
The process starts by the generation of a supertriangle that encompasses all the points
in $S$. At the end, all triangles that share one edge with the supertriangle
are removed. The addition of one point is illustrated in figure \ref{incremental_algo}

To add one point $P$, all the triangles whose circumcircle contains $P$ are first
removed. The resulting hole in the triangulation has a polygonal shape. New triangles
are formed between $P$ and the outside edges of the polygon.

\end{document}